\documentclass[a4paper,12pt]{article}

\usepackage[top=1.2in,right=0.5in,left=.5in,bottom=1.2in]{geometry}

\usepackage{latexsym}
\usepackage{amssymb}
\usepackage{float}
\usepackage{graphics,graphicx}
\usepackage{tabularx}
\usepackage{amsfonts}
\usepackage{amsmath}
\usepackage{enumerate}
\usepackage{booktabs}
\usepackage{multirow}
\usepackage{sectsty}
\usepackage[affil-it,auth-sc]{authblk}
\usepackage{subfig}
\usepackage{color}
\usepackage{mathtools}
\usepackage{mathrsfs}
\usepackage{bmpsize}
\usepackage{xcolor}
\usepackage{siunitx}

\sectionfont{\normalsize}
\subsectionfont{\normalsize}
\usepackage{fancyhdr}            
\fancypagestyle{plain}{%
  \fancyhead{}                                     
  \fancyfoot[CE,CO]{\thepage}       
  \fancyhead[CE,CO]{\textcolor[rgb]{0.64,0.15,0.15}{Published in \emph{International Journal of Solids and Structures} (2019) 
  \\
doi: http://dx.doi.org/10.1016/j.ijsolstr.2018.12.005}}
\setlength{\headheight}{14.5pt}}

\newcommand{\footmsg}[1]{%
  \let\temp\thempfn%
  \def\thempfs{}
  \footnotetext{#1}
  \let\tempfn\temp}

\begin{document}

\newcommand{\singlespace}{\baselineskip=12pt\lineskiplimit=0pt\lineskip=0pt}
\def\ds{\displaystyle}

\newcommand{\beq}{\begin{equation}}
\newcommand{\eeq}{\end{equation}}
\newcommand{\lb}{\label}
\newcommand{\ph}{\phantom}
\newcommand{\beqar}{\begin{eqnarray}}
\newcommand{\eeqar}{\end{eqnarray}}
\newcommand{\barr}{\begin{array}}
\newcommand{\earr}{\end{array}}
\newcommand{\jump}{\parallel}
\newcommand{\Ehat}{\hat{E}}
\newcommand{\That}{\hat{\bf T}}
\newcommand{\Ahat}{\hat{A}}
\newcommand{\chat}{\hat{c}}
\newcommand{\shat}{\hat{s}}
\newcommand{\khat}{\hat{k}}
\newcommand{\muhat}{\hat{\mu}}
\newcommand{\mc}{M^{\scriptscriptstyle C}}
\newcommand{\mei}{M^{\scriptscriptstyle M,EI}}
\newcommand{\mec}{M^{\scriptscriptstyle M,EC}}
\newcommand{\hbeta}{{\hat{\beta}}}
\newcommand{\rec}[2]{\left( #1 #2 \ds{\frac{1}{#1}}\right)}
\newcommand{\rep}[2]{\left( {#1}^2 #2 \ds{\frac{1}{{#1}^2}}\right)}
\newcommand{\derp}[2]{\ds{\frac {\partial #1}{\partial #2}}}
\newcommand{\derpn}[3]{\ds{\frac {\partial^{#3}#1}{\partial #2^{#3}}}}
\newcommand{\dert}[2]{\ds{\frac {d #1}{d #2}}}
\newcommand{\dertn}[3]{\ds{\frac {d^{#3} #1}{d #2^{#3}}}}
\newcommand{\ct}{\captionof{table}}
\newcommand{\cf}{\captionof{figure}}

\def\c{{\circ}}
\def\bob{{\, \underline{\overline{\otimes}} \,}}
\def\ob{{\, \underline{\otimes} \,}}
\def\scalp{\mbox{\boldmath$\, \cdot \, $}}
\def\gdp{\makebox{\raisebox{-.215ex}{$\Box$}\hspace{-.778em}$\times$}}
\def\daa{\makebox{\raisebox{-.050ex}{$-$}\hspace{-.550em}$: ~$}}
\def\mK{\mbox{${\mathcal{K}}$}}
\def\cK{\mbox{${\mathbb {K}}$}}

\def\Xint#1{\mathchoice
   {\XXint\displaystyle\textstyle{#1}}%
   {\XXint\textstyle\scriptstyle{#1}}%
   {\XXint\scriptstyle\scriptscriptstyle{#1}}%
   {\XXint\scriptscriptstyle\scriptscriptstyle{#1}}%
   \!\int}
\def\XXint#1#2#3{{\setbox0=\hbox{$#1{#2#3}{\int}$}
     \vcenter{\hbox{$#2#3$}}\kern-.5\wd0}}
\def\ddashint{\Xint=}
\def\fpint{\Xint=}
\def\dashint{\Xint-}
\def\cpvint{\Xint-}
\def\intl{\int\limits}
\def\cpvintl{\cpvint\limits}
\def\fpintl{\fpint\limits}
\def\ointl{\oint\limits}
\def\bA{{\bf A}}
\def\ba{{\bf a}}
\def\bB{{\bf B}}
\def\bb{{\bf b}}
\def\bc{{\bf c}}
\def\bC{{\bf C}}
\def\bD{{\bf D}}
\def\bE{{\bf E}}
\def\be{{\bf e}}
\def\bbf{{\bf f}}
\def\bF{{\bf F}}
\def\bG{{\bf G}}
\def\bg{{\bf g}}
\def\bi{{\bf i}}
\def\bH{{\bf H}}
\def\bK{{\bf K}}
\def\bL{{\bf L}}
\def\bM{{\bf M}}
\def\bN{{\bf N}}
\def\bn{{\bf n}}
\def\bm{{\bf m}}
\def\b0{{\bf 0}}
\def\bo{{\bf o}}
\def\bX{{\bf X}}
\def\bx{{\bf x}}
\def\bP{{\bf P}}
\def\bp{{\bf p}}
\def\bQ{{\bf Q}}
\def\bq{{\bf q}}
\def\bR{{\bf R}}
\def\bS{{\bf S}}
\def\bs{{\bf s}}
\def\bT{{\bf T}}
\def\bt{{\bf t}}
\def\bU{{\bf U}}
\def\bu{{\bf u}}
\def\bv{{\bf v}}
\def\bw{{\bf w}}
\def\bW{{\bf W}}
\def\by{{\bf y}}
\def\bz{{\bf z}}
\def\T{{\bf T}}
\def\Te{\textrm{T}}
\def\Id{{\bf I}}
\def\bxi{\mbox{\boldmath${\xi}$}}
\def\balpha{\mbox{\boldmath${\alpha}$}}
\def\bbeta{\mbox{\boldmath${\beta}$}}
\def\bepsilon{\mbox{\boldmath${\epsilon}$}}
\def\bvarepsilon{\mbox{\boldmath${\varepsilon}$}}
\def\bomega{\mbox{\boldmath${\omega}$}}
\def\bphi{\mbox{\boldmath${\phi}$}}
\def\bsigma{\mbox{\boldmath${\sigma}$}}
\def\bfeta{\mbox{\boldmath${\eta}$}}
\def\bDelta{\mbox{\boldmath${\Delta}$}}
\def\btau{\mbox{\boldmath $\tau$}}
\def\tr{{\rm tr}}
\def\dev{{\rm dev}}
\def\div{{\rm div}}
\def\Div{{\rm Div}}
\def\Grad{{\rm Grad}}
\def\grad{{\rm grad}}
\def\Lin{{\rm Lin}}
\def\Sym{{\rm Sym}}
\def\Skw{{\rm Skew}}
\def\abs{{\rm abs}}
\def\Re{{\rm Re}}
\def\Im{{\rm Im}}
\def\capB{\mbox{\boldmath${\mathsf B}$}}
\def\capC{\mbox{\boldmath${\mathsf C}$}}
\def\capD{\mbox{\boldmath${\mathsf D}$}}
\def\capE{\mbox{\boldmath${\mathsf E}$}}
\def\capG{\mbox{\boldmath${\mathsf G}$}}
\def\tcapG{\tilde{\capG}}
\def\capH{\mbox{\boldmath${\mathsf H}$}}
\def\capK{\mbox{\boldmath${\mathsf K}$}}
\def\capL{\mbox{\boldmath${\mathsf L}$}}
\def\capM{\mbox{\boldmath${\mathsf M}$}}
\def\capR{\mbox{\boldmath${\mathsf R}$}}
\def\capW{\mbox{\boldmath${\mathsf W}$}}

\def\i{\mbox{${\mathrm i}$}}
\def\mC{\mbox{\boldmath${\mathcal C}$}}
\def\mB{\mbox{${\mathcal B}$}}
\def\mE{\mbox{${\mathcal{E}}$}}
\def\mL{\mbox{${\mathcal{L}}$}}
\def\mK{\mbox{${\mathcal{K}}$}}
\def\mV{\mbox{${\mathcal{V}}$}}
\def\C{\mbox{\boldmath${\mathcal C}$}}
\def\E{\mbox{\boldmath${\mathcal E}$}}

\def\AAM{{\it Advances in Applied Mechanics }}
\def\ACME{{\it Arch. Comput. Meth. Engng.}}
\def\ARMA{{\it Arch. Rat. Mech. Analysis}}
\def\AMR{{\it Appl. Mech. Rev.}}
\def\ASCEEM{{\it ASCE J. Eng. Mech.}}
\def\ACTA{{\it Acta Mater.}}
\def\CMAME {{\it Comput. Meth. Appl. Mech. Engrg.}}
\def\CRAS{{\it C. R. Acad. Sci. Paris}}
\def\CRM{{\it Comptes Rendus M\'ecanique}}
\def\EFM{{\it Eng. Fracture Mechanics}}
\def\EJMA{{\it Eur.~J.~Mechanics-A/Solids}}
\def\IJES{{\it Int. J. Eng. Sci.}}
\def\IJF{{\it Int. J. Fracture}}
\def\IJMS{{\it Int. J. Mech. Sci.}}
\def\IJNAMG{{\it Int. J. Numer. Anal. Meth. Geomech.}}
\def\IJP{{\it Int. J. Plasticity}}
\def\IJSS{{\it Int. J. Solids Structures}}
\def\IngA{{\it Ing. Archiv}}
\def\JAM{{\it J. Appl. Mech.}}
\def\JAP{{\it J. Appl. Phys.}}
\def\JAE{{\it J. Aerospace Eng.}}
\def\JE{{\it J. Elasticity}}
\def\JM{{\it J. de M\'ecanique}}
\def\JMPS{{\it J. Mech. Phys. Solids}}
\def\JSV{{\it J. Sound and Vibration}}
\def\MACRO{{\it Macromolecules}}
\def\MMT{{\it Mech. Mach. Th.}}
\def\MOM{{\it Mech. Materials}}
\def\MMS{{\it Math. Mech. Solids}}
\def\MMT{{\it Metall. Mater. Trans. A}}
\def\MPCPS{{\it Math. Proc. Camb. Phil. Soc.}}
\def\MSE{{\it Mater. Sci. Eng.}}
\def\NATURE{{\it Nature}}
\def\NATUREM{{\it Nature Mater.}}
\def\PHIL{{\it Phil. Trans. R. Soc.}}
\def\PMPS{{\it Proc. Math. Phys. Soc.}}
\def\PNAS{{\it Proc. Nat. Acad. Sci.}}
\def\PRE{{\it Phys. Rev. E}}
\def\PRL{{\it Phys. Rev. Letters}}
\def\PRSL{{\it Proc. R. Soc.}}
\def\RIIT{{\it Rozprawy Inzynierskie - Engineering Transactions}}
\def\ROCK{{\it Rock Mech. and Rock Eng.}}
\def\QAM{{\it Quart. Appl. Math.}}
\def\QJMAM{{\it Quart. J. Mech. Appl. Math.}}
\def\SCIENCE{{\it Science}}
\def\SCRMAT{{\it Scripta Mater.}}
\def\SM{{\it Scripta Metall.}}
\def\ZAMM{{\it Z. Angew. Math. Mech.}}
\def\ZAMP{{\it Z. Angew. Math. Phys.}}
\def\ZVDI{{\it Z. Verein. Deut. Ing.}}

\def\salto#1#2{
[\mbox{\hspace{-#1em}}[#2]\mbox{\hspace{-#1em}}]}

\renewcommand\Affilfont{\itshape}
\setlength{\affilsep}{1em}
\renewcommand\Authsep{, }
\renewcommand\Authand{ and }
\renewcommand\Authands{ and }
\setcounter{Maxaffil}{2}

\title{Snapping of elastic strips with controlled ends}

\author{Alessandro Cazzolli, Francesco Dal Corso}
 \affil[]{DICAM, University of Trento, via~Mesiano~77, I-38123
Trento, Italy.}

\date{}
\maketitle \footnotetext[1]{Corresponding author: Francesco Dal
Corso fax: +39 0461 282599; tel.: +39 0461 282522; web-site:
http://www.ing.unitn.it/$\sim$dalcorsf/; e-mail:
francesco.dalcorso@unitn.it}

\date{}
\maketitle

\begin{abstract}
Snapping mechanisms are investigated for an elastic strip with ends imposed to move and rotate in time.
Attacking  the problem analytically via Euler's elastica and the second variation of the total potential energy,
the number of stable equilibrium configurations is disclosed by varying the kinematics of the strip ends.
This result leads to the definition of a \lq universal snap surface', collecting the sets of critical boundary conditions for which
the system snaps. The elastic energy release at snapping is also investigated, providing useful insights for the optimization of impulsive motion.
The theoretical predictions are finally
 validated through comparisons with experimental results
and finite element simulations, both fully confirming the reliability of the introduced universal surface.
The presented analysis may find applications in a wide range of technological fields, as for instance energy harvesting and jumping robots.
\end{abstract}


\section{Introduction}

Snap instability is a well-known phenomenon in  mechanics for which a structural system suffers
a sudden and dramatic change in the deformed configuration triggered by a small variation in the loading conditions.
This behaviour is explained as the consequence of the stability loss for the deformed configuration, so that the structure   dynamically moves towards a non-adjacent configuration
through a partial release of its elastic energy. Classical examples of snap mechanisms can be found in shell structures, also in everyday life,
as for instance when squeezing an empty can of soda.
Other examples may be found in nature, as in the
case of click-beetles (elateridae) \cite{evanselaterid1}, insects able to turn on their side when initially lying on their back by means
of a jump realized by a snap mechanism.

Following the new paradigm of exploiting (instead of avoiding) instabilities in the structural design for
advanced applications \cite{hu, reis}, in the last years  many researchers  have investigated the snap mechanisms
towards the realization of bistable or multistable devices \cite{armanini,camescasse,camescasse2,cazotte,chen2008,chen2012,restrepo,
schioler},
metamaterials \cite{frazier, nadkarni, raney},  locomotion \cite{mochiyama, tsuda, yamada, yang},
 and energy harvesting \cite{harne, kimalcubo}.
Because the typical approach adopted in these investigations is to focus on specific evolutive mechanical problems and specific structural properties,
the evaluation of the whole set of critical snap conditions is missed from a general perspective.

The aim of the present study is to provide a general criterion to be exploited in the nonlinear design of structures to obtain (or to avoid)
snap mechanisms.
With reference to the model of an inextensible (weightless) linearly elastic strip,
the number of stable equilibrium configurations is disclosed by varying the parameters defining the kinematics of the strip ends, which
are reduced to a normalized  distance between the two ends and the two rotations of the ends.
The stability of the equilibrium configurations, expressed in closed form through elliptic integrals,
is assessed analyzing the sign of the second variation of the total potential energy at fixed ends conditions.
This analysis allows us to define the correspondence between
the ends' kinematics and the presence  of multiple stable configurations, so that the set of boundary conditions for which one of these configurations loses
its stability follows.

The obtained results are used to define universal snap surfaces, collecting the whole sets of critical boundary conditions for which the system snaps.
The energy release at snapping is also estimated and investigated by means of a dimensionless analysis for varying the snap conditions.
The theoretical predictions, obtained analytically within a quasi-static framework, are experimentally validated through comparisons with data available for specific symmetric boundary conditions
and observations on a physical model proposed for investigating non-symmetric boundary conditions. Finally, finite element simulations performed with ABAQUS show the reliability of
the presented universal surface in the case of evolutions with moderate velocity and its limits in the case of very fast ends evolutions.

\section[Governing equations and  the planar \emph{elastica}]{Governing equations and  the planar \emph{elastica}}

\begin{figure}[!htb]
\begin{center}
\includegraphics[width=160mm]{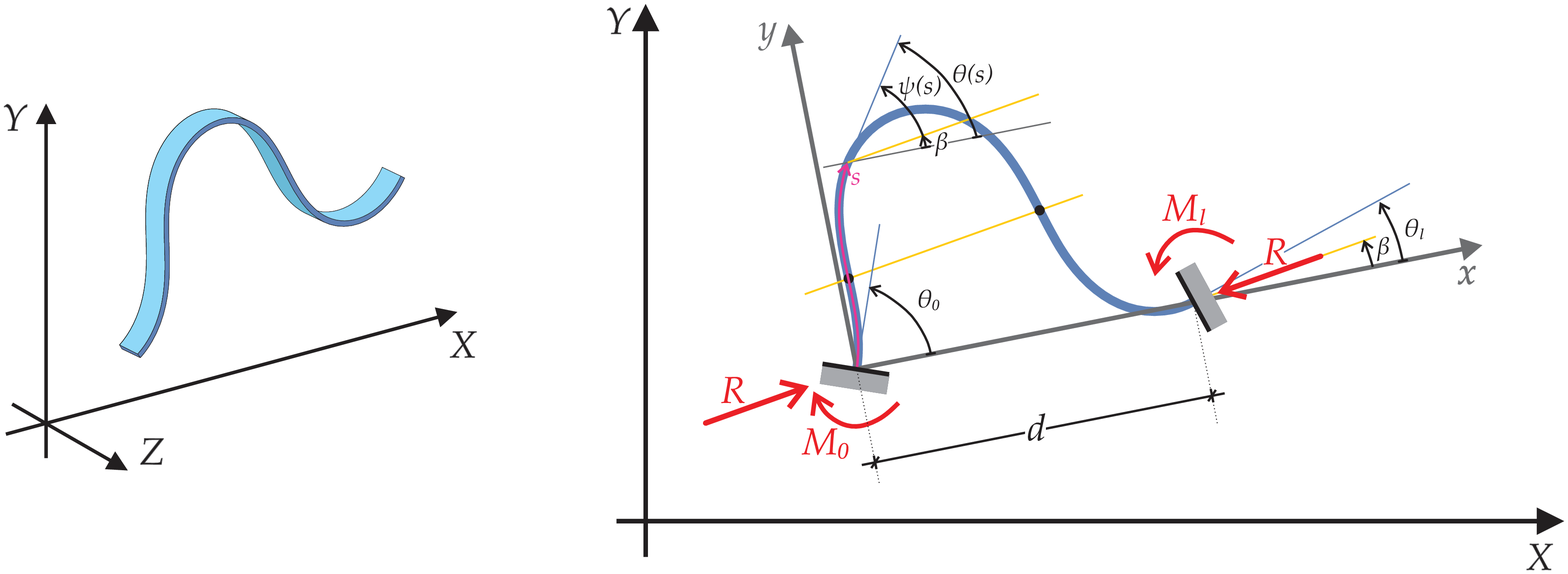}
\end{center}
\caption{\footnotesize
Generic deformed configuration for a strip of length $l$ with kinematically controlled ends within the absolute reference system $X-Y$ and the relative reference system $x-y$. Neglecting rigid-body motions, the equilibrium configuration is dependent only on the three independent kinematic quantities $d, \theta_0,$ and $\theta_l$. Reaction forces and moments at the two controlled ends are also reported. 
}
\label{fig_beamCD}
\end{figure}

An inextensible elastic strip of length $l$ is considered to be deformed within a plane orthogonal to the axis of minimum momenta of inertia of the strip's cross section.
The strip has a uniform cross section and is initially flat, so that its centerline is described by a straight line in the undeformed configuration.
Neglecting the effects of self weight and disregarding rigid body motions, the mechanical fields along the generic curvilinear coordinate $s\in[0,l]$ can be represented in the local reference system $x-y$ (Fig. \ref{fig_beamCD}), with origin  at one strip's end ($s=0$), and $x$-axis passing by the other one ($s=l$), so that $x(0)=y(0)=y(l)=0$, and pointing from the initial to the final curvilinear coordinate. The primary kinematic field is the rotation angle $\theta(s)$,
which measures the rotation of the structure's centerline  with respect to the $x$-axis, from which, considering the inextensibility constraint, the position fields can be obtained as
\begin{equation}
\label{BCkinematic}
x(s)=\int_{0}^{s}{\cos{\theta(\varsigma)}}\, \textup{d}\varsigma, \qquad
y(s)=\int_{0}^{s}{\sin{\theta(\varsigma)}}\, \textup{d}\varsigma,
\end{equation}
so that the condition $y(l)=0$ implies the following isoperimetric constraint on the rotation field $\theta(s)$
\beq\label{BCkinematicdef}
\int_{0}^{l}{\sin{\theta(s)}}\, \textup{d}s=0.
\eeq

The strip is considered subject to kinematic boundary conditions in terms of position and rotation at both ends, which slowly move in time along \emph{quasi-static evolutions}. 
Considering the $x-y$ reference system allows for the rational interpretation of the boundary conditions imposed at the two rod's ends. Indeed, the six kinematic boundary conditions (two positions and one rotation at each end)  affect the strip configuration by means of the following three primary kinematical  quantities
\beq\label{BCkinematics}
x(l)=d,\qquad
\theta(0)=\theta_0,\qquad
\theta(l)=\theta_l,
\eeq
being $d$ the distance between the two clamps, $d\in[0,l]$ where the lower bound is given by the definition of the x-axis direction while the upper bound is related to the inextensibility assumption (although extensibility may affect the mechanical response even in the proximity of the limit condition,  $d\simeq l$).

The planar behaviour of the considered strip is modelled as the Euler elastica so that the
bending moment $M$ is given by $M(s)=B \theta'(s)$ where the symbol $'$ stands for the derivative with respect to the curvilinear coordinate $s$ and
 $B$ is the bending stiffness, constant because the strip's cross section is uniform. Considering that the deformed configuration is described by imposing the distance $d$ and the two rotations $\theta_0$ and $\theta_l$, the total potential energy $\mathcal{V}$  under \emph{quasi-static conditions} is given by
\begin{equation}
\label{eptSP}
\begin{split}
\mathcal{V}\big(\theta(s)\big)=&\,\mathcal{E}+R_x\left[-d+\int_{0}^{l}\cos{\theta(s)} \,\textup{d}s\right]
+R_y\left[\int_{0}^{l}\sin{\theta(s)}\,\textup{d}s\right]+M_l\bigg[\theta_l-\theta(l)\bigg]-M_0\bigg[\theta_0-\theta(0)\bigg],
\end{split}
\end{equation}
where $\mathcal{E}$ is the elastic energy stored within the strip
\beq\label{energydefinition}
\mathcal{E}=\frac{B}{2}\int_{0}^{l}{\theta'(s)}^2\,\textup{d}s,
\eeq
while the quantities $R_x$ and $R_y$ are the Lagrangian multipliers representing the reaction forces components at both ends along the $x$ and $y$ directions and, similarly, $M_l$ and $M_0$ are those referred to the rotational degrees of freedom of the strip's coordinates $s=l$ and $s=0$.

The annihilation of the first variation of the functional $\mathcal{V}$ leads to the differential equation of the Euler elastica \cite{bigonidef}
\begin{equation}
\label{diffpsi}
 \psi''(s)+\gamma^2\sin{\psi(s)}=0  \qquad  \forall s \in [0 , l].
\end{equation}
being $\psi(s)=\theta(s)-\beta$ the auxiliary angle,  $\gamma^2=R/B$ the normalized load parameter, $R=\sqrt{R_x^2+R_y^2}$ the total reaction force and $\beta$ its inclination with respect to the $x$-axis (Fig \ref{fig_beamCD}).

A first integration of the equation ($\ref{diffpsi}$) leads to
\begin{equation}
\label{psiconstnoflex}
 \psi'(s)=\pm\gamma\sqrt{2}\sqrt{\cos{\psi(s)}+\Upsilon},
\end{equation}
where $\Upsilon$  is the constant of integration to be evaluated with reference to
the boundary conditions and defining the elastica as a part of a \textit{non-inflectional mother curve} ($\Upsilon\in[1,\infty]$)
or an \textit{inflectional mother curve} ($\Upsilon\in[-1,1]$) \cite{love}.

In the following paragraphs, the analytical description of the deformed configurations is presented with reference to the number $m\in \mathbb{N}^0$ of  \lq inflection points',
corresponding to the curvilinear coordinates within the set $s\in(0,l)$ where the curvature vanishes. It is remarked that points with null curvature located at both ends are not considered in the definition of $m$.
For convenience, the constant of integration $\Upsilon$ is  defined distinguishing the fundamental cases of the absence ($m=0$) and the presence ($m\neq0$) of inflection points
and the deformed configurations are described, respectively, in terms of the equations related to the so-called  \textit{non-inflectional} and \textit{inflectional elastica}.
Therefore it is worth to remark that, differently from Love \cite{love}, the deformed configurations with $m=0$ associated with an \textit{inflectional mother curve}
are here described  for simplicity through the expressions used for the \textit{non-inflectional mother curve} (for which the ideal unlimited elastica, $-\infty < s < \infty$, has no inflection points)
but restricted to the physical range of the curvilinear coordinate, $s\in[0,l]$ (as shown in Sect. \ref{elasticae}).

\paragraph{Absence of inflection points along the strip  ($m=0$).}
In this case $\psi'(s)\neq0$ for $s\in(0,l)$ and the integration constant $\Upsilon$
can be defined as a function of an unknown parameter $\xi$ as
\beq\label{primaEQ}
\Upsilon=\frac{2-\xi^2}{\xi^2},\qquad
\xi\in\left[0,\,\sqrt{\frac{2}{1-\ds\min_{s\in(0,l)}\left\{\cos\psi(s)\right\}}}\right],
\eeq
where the upper bound for the parameter $\xi$ restricts the solution at equilibrium to non-null curvature values within the set $s\in(0,l)$.
It is noted that, differently from Love \cite{love}, the range of $\xi$ is extended to values higher than 1, which are related to inflectional mother curves.

Furthermore, an auxiliary rotation field is introduced as $\chi(s)=\psi(s)/2$ whose values at the two ends are defined as $\chi(0)=\chi_0$ and $\chi(l)=\chi_l$, given by
\begin{equation}
\label{chi0l}
\begin{split}
\chi_0=\frac{\theta_0-\beta}{2},\qquad
\chi_{l}=\frac{\theta_l-\beta}{2}.
\end{split}
\end{equation}

\paragraph{Presence  of $m$ inflection points  along the strip  ($m\neq 0$).}

In this case $\psi'(\widehat{s}_j)=0$ at the curvilinear coordinates $\widehat{s}_j\in(0,l)$ with $j\in[1,m]$ ordering the inflection points with respect to their curvilinear coordinate,
$\widehat{s}_{j}<\widehat{s}_{j+1}$ . Eqn (\ref{psiconstnoflex}) implies that
the integration constant $\Upsilon$ is a function of the angle $\psi(\widehat{s}_1)=\widehat{\psi}_1$ measured at the   inflection point  along the strip 
and closest to  the origin as
\beq
\Upsilon=-\cos{\widehat{\psi}_1}.
\eeq
Being $\cos\widehat{\psi}_j=\cos{\widehat{\psi}_1}$ with $\psi(\widehat{s}_j)=\widehat{\psi}_j$ ($j=1,...,m$),
 the rotation angle at the  $j$-th inflection point is given by
\beq
\theta(\widehat{s}_j)=-(-1)^j\theta(\widehat{s}_1)+\left[1+(-1)^j\right]\beta.
\eeq
Because the curvature changes sign at each inflection point,
the solution (\ref{psiconstnoflex}) for the curvature can be rewritten as
\begin{equation}
\psi'(s)=
\left\{
\begin{array}{lll}
&(-1)^p\gamma\sqrt{2}\sqrt{\cos{\psi(s)}-\cos{\widehat{\psi}_1}},\qquad\qquad 
&\forall s\in[0,\,\widehat{s}_{1}],\\[3mm]
&(-1)^p(-1)^{j}\gamma\sqrt{2}\sqrt{\cos{\psi(s)}-\cos{\widehat{\psi}_1}},\qquad\qquad 
&\forall s\in[\widehat{s}_j,\,\widehat{s}_{j+1}]\qquad \forall\,\, j\in[1,m-1],\\[3mm]
&(-1)^p(-1)^{m}\gamma\sqrt{2}\sqrt{\cos{\psi(s)}-\cos{\widehat{\psi}_1}},\qquad \qquad 
&\forall s\in[\widehat{s}_m,\,l],
\end{array}\right.
\end{equation}
where $p$ is a boolean parameter defining the curvature sign at the left end $s=0$, namely
$p=0$ ($p=1$) when the curvature is positive (negative) at the left end, $\theta'(s=0)>0$  ($\theta'(s=0)<0$).\footnote{In the case of null curvature at the initial coordinate, $\theta'(s=0)=0$, the boolean $p$ is defined by the sign of the curvature at positive infinitesimal values of the coordinate, $\theta'(s=0^+)$.
}
Towards the solution achievement it is instrumental to introduce the parameter $\eta\in \,[0,1]$ and the auxiliary field $\omega(s)$ defined as
\begin{equation}
\label{changev} \eta=\sin{\frac{\hat{\psi}_1}{2}}, \qquad
\eta\sin{\omega(s)}=\sin{\frac{\psi(s)}{2}}.
\end{equation}
The values of the function $\omega(s)$ attained at the two ends, $\omega(s=0)=\omega_0$ and $\omega(s=l)=\omega_l$, represent the fundamental parameters
involved in the problem resolution and depend on the parameters $\beta$ and $\eta$ as follows:
\begin{equation}
\omega_0=\arcsin{\left(\dfrac{1}{\eta}\sin{\dfrac{\theta_{0}-\beta}{2}}\right)},\qquad
\omega_{l}=(-1)^m\,\arcsin{\left(\dfrac{1}{\eta}\sin{\dfrac{\theta_{l}-\beta}{2}}\right)}+(-1)^p\,m\pi.
\end{equation}

\section{The \emph{elasticae} joining  two constrained ends}\label{elasticae}

A further integration of the differential equation (\ref{psiconstnoflex}) leads to the following relations for the normalized load $\gamma$
\begin{equation}
\label{gammasol}
\gamma l= \left\{
\begin{array}{lll}
\xi\left|K(\chi_{l},\,\xi)-K(\chi_{0},\,\xi)\right|,\qquad\qquad&m=0,\\[3mm]
\left|K(\omega_{l},\,\eta)-K(\omega_0,\,\eta)\right|,\qquad\qquad &m\neq 0,
\end{array}
\right.
\end{equation}
and the rotation field
\begin{equation}\label{rotasol}
\theta(s)= \left\{
\begin{array}{lll}
\beta+2\,\textup{am}\left(\dfrac{s}{l}\left(K(\chi_{l},\,\xi)-K(\chi_{0},\,\xi)\right)+K(\chi_{0},\,\xi),\,\xi\right), \qquad\qquad & m=0,\\[3mm]
\beta+2\,\arcsin{\left[\eta\,\textup{sn}\left(\dfrac{s}{l}\left(K(\omega_{l},\,\eta)-K(\omega_0,\,\eta)\right)+K(\omega_0,\,\eta),\,\eta\right)\right]},\qquad\qquad & m\neq 0.
\end{array}
\right.
\end{equation}
In equations (\ref{gammasol}) and (\ref{rotasol}), $K$ is the \textit{Jacobi's incomplete elliptic integral of the first kind},
$\textup{am}$ is the \textit{Jacobi's amplitude function},
$\textup{sn}$  is the \textit{Jacobi's sine amplitude function},
\begin{equation}
 K\left(\sigma,\,\varphi\right)=\int_{0}^{\sigma}\frac{\textup{d}\phi}{\sqrt{1-\varphi^2\sin^2{\phi}}},
 \qquad
 \sigma=\textup{am}\bigg(K\left(\sigma,\,\varphi\right),\,\varphi\bigg),
 \qquad
 \textup{sn}(u,\,\varphi)=\sin{\left(\textup{am}(u,\,\varphi)\right)}.
\end{equation}

It is remarked that, when the  deformed configuration is associated with an inflectional mother curve, $\Upsilon\in[-1,1]$, but there are no inflection points  along the strip, $m=0$,
 the analytical representation of the normalized load $\gamma$, eqn (\ref{gammasol}), and of  the rotation field $\theta(s)$, eqn (\ref{rotasol}),
 for the case $m=0$ coincides with that for the case $m\neq 0$.\footnote{In addition to the
general property for the reciprocal modulus transformation for the Jacobi's sine amplitude function (Byrd \cite{byrd}, his eqn 162.01, pag 38)
\beq
\textup{sn}\left(\sigma \varphi,\frac{1}{\varphi}\right)=\varphi\,\,\textup{sn}\left(\sigma,\varphi\right),
\eeq
under the circumstance $\Upsilon\in[-1,1]$ and $m=0$ the following properties hold:
\beq
\xi=1/\eta, \qquad
\sin\omega(s)=\xi\sin\chi(s),\qquad
K\left(\omega(s),\,\eta\right)=\xi K\left(\chi(s),\,\xi\right).
\eeq
}

From the rotation field (\ref{rotasol}), the elastic energy $\mathcal{E}$, eqn(\ref{energydefinition}), is given by
\beq
\label{energyexplicit}
\mathcal{E}= \left\{
\begin{array}{lll}
\ds\frac{2B}{l}\big[K(\chi_{l},\,\xi)-K(\chi_{0},\,\xi)\big]\big[E(\chi_{l},\,\xi)-E(\chi_{0},\,\xi)\big], \qquad & m=0,\\[3mm]
\ds\frac{2B}{l}\big[K(\omega_{l},\,\eta)-K(\omega_{0},\,\eta)\big]\bigg\{E(\omega_{l},\,\eta)-E(\omega_{0},\,\eta)-(1-\eta^2)[K(\omega_{l},\,\eta)-K(\omega_{0},\,\eta)\big]\bigg\},\qquad
& m\neq 0,
\end{array}
\right.
\eeq
and the deformed shape at equilibrium can be evaluated from the position field (\ref{BCkinematic}) and expressed as
\begin{equation}
\label{X1X2}
x(s)=\big[\cos{\beta}\,\mathcal{A}_{m}(s)+\sin{\beta}\,\mathcal{B}_{m}(s)\big]l,\qquad
 y(s)=\big[\sin{\beta}\,\mathcal{A}_{m}(s)-\cos{\beta}\,\mathcal{B}_{m}(s)\big]l
\end{equation}
with the functions $\mathcal{A}_{m}(s)$ and $\mathcal{B}_{m}(s)$ given by
\begin{subequations}\label{ABAB}
 \begin{align}
\mathcal{A}_{m}(s)= \left\{
\begin{array}{lll}
\ds\frac{2}{\xi^2}\frac{\mathscr{E}\left(\dfrac{s}{l}\left(K(\chi_{l},\,\xi)-K(\chi_{0},\,\xi)\right)
+K(\chi_{0},\,\xi),\,\xi\right)+\mathscr{E}\left(K(\chi_{0},\,\xi),\,\xi\right)}{K(\chi_{l},\,\xi)-K(\chi_{0},\,\xi)}
-\frac{2-\xi^2}{\xi^2}\frac{s}{l},\qquad &m=0, \\[5mm]
\ds2\frac{\mathscr{E}\left(\dfrac{s}{l}\left(K(\omega_{l},\,\eta)-K(\omega_0,\,\eta)\right)+K(\omega_0,\,\eta),\,\eta\right)
-\mathscr{E}\left(K(\omega_0,\,\eta),\,\eta\right)}{K(\omega_{l},\,\eta)-K(\omega_0,\,\eta)}
-\frac{s}{l},\qquad & m\neq 0,
\end{array}
\right.
\end{align}
\begin{align}
\mathcal{B}_{m}(s)= \left\{
\begin{array}{lll}
\ds\frac{2}{\xi^2}\frac{\textup{dn}\left(\dfrac{s}{l}\left(K(\chi_{l},\,\xi)-K(\chi_{0},\,\xi)\right)+K(\chi_{0},\,\xi),\,\xi\right)
-\textup{dn}\left(K(\chi_{0},\,\xi),\,\xi\right)}{K(\chi_{l},\,\xi)-K(\chi_{0},\,\xi)},\qquad &m=0,
\\[5mm]
\ds 2\,\eta\,\frac{\textup{cn}\left(\dfrac{s}{l}\left(K(\omega_{l},\,\eta)-K(\omega_0,\,\eta)\right)+K(\omega_0,\,\eta),\,\eta\right)
-\textup{cn}\left(K(\omega_0,\,\eta),\,\eta\right)}{K(\omega_{l},\,\eta)-K(\omega_0,\,\eta)},\qquad & m\neq 0.
\end{array}
\right.
\end{align}
\end{subequations}
In eqns (\ref{ABAB}) the function $\textup{cn}$ is the \textit{Jacobi's cosine amplitude function},
$\mathscr{E}$ is the  \textit{Jacobi's epsilon function}, and $ \textup{dn}$ is the \textit{Jacobi's elliptic function},
defined as
\begin{equation}
 \textup{cn}(u,\,\varphi)=\cos{\left(\textup{am}(u,\,\varphi)\right)},\qquad
 \mathscr{E}\left(\sigma,\,\varphi\right)=E(\textup{am}(\sigma,\,\varphi),\,\varphi),\qquad
  \textup{dn}(u,\,\varphi)=\sqrt{1-\varphi^2\,\textup{sn}^2(u,\,\varphi)},
\end{equation}
and $E$ is the \textit{Jacobi's incomplete elliptic integral of
the second kind}
\begin{equation}
 E\left(\sigma,\,\varphi\right)=\int_{0}^{\sigma}\sqrt{1-\varphi^2\sin^2{\phi}}\,\textup{d}\phi.
\end{equation}

Imposing  specific boundary conditions at the strip's ends, the \emph{elastica} joining the two constrained ends can be found.
The two isoperimetric constraints, $x(l)=d$ and $y(l)=0$, are enforced through the following nonlinear system:
\begin{equation}
\label{sistPOSnoflex}
 \left\{
  \begin{aligned}
& d\cos{\beta}= \mathcal{A}_m(l) l,\\
& d\sin{\beta}= \mathcal{B}_m(l) l,
  \end{aligned}
\right.
\end{equation}
where, from equation (\ref{ABAB}), the quantities $\mathcal{A}_{m}(l)$ and $\mathcal{B}_{m}(l)$ are given by
\begin{equation}
\resizebox{\textwidth}{!}{$
\mathcal{A}_{m}(l)= \left\{
\begin{array}{lll}
\ds 1-\frac{2}{\xi^2}\left(1-\frac{E(\chi_{l},\,\xi)-E(\chi_{0},\,\xi)}{K(\chi_{l},\,\xi)-K(\chi_{0},\,\xi)}\right),
\\[5mm]
\ds 2\frac{E(\omega_{l},\,\eta) -E(\omega_0,\,\eta)}{K(\omega_{l},\,\eta) -K(\omega_0,\,\eta)}-1,
\end{array}
\right.
\,
\mathcal{B}_{m}(l)= \left\{
\begin{array}{lll}
\ds \frac{2}{\xi^2}\frac{\sqrt{1-\xi^2\sin^2{\chi_{l}}}-\sqrt{1-\xi^2\sin^2{\chi_{0}}}}{K(\chi_{l},\,\xi)-K(\chi_{0},\,\xi)}, \quad &m=0,
\\[5mm]
\ds \frac{2\eta\left(\cos{\omega_{l}}-\cos{\omega_0}\right)}{K(\omega_{l},\,\eta)-K(\omega_0,\,\eta)},\quad & m\neq 0.
\end{array}
\right.
$}
\end{equation}

The nonlinearity of the system (\ref{sistPOSnoflex}) leads to possible non-uniqueness for the equilibrium configuration,
providing interesting conditions of bifurcation or snap instability during a quasi-static variation of the boundary conditions as shown in the next Section.

As a final remark,  equations (\ref{primaEQ})-(\ref{ABAB})
can be directly exploited to  attack various boundary value problems by properly imposing the nonlinear system to be solved, so that the present formulation can also be considered to treat structural systems as an alternative to 
previously adopted procedures \cite{zhang}.

\subsection{Stability of the elastica with constrained ends}\label{stability}
Extending previously presented procedures \cite{batista},\cite{bigoni},  \cite{lev1}, the stability of the achieved equilibrium configurations is judged with reference to small and compatible perturbations in the rotation field $\theta(s)$.
In particular, the stability is connected to the sign of the second variation of the total potential
energy (\ref{eptSP}), which referring to a weak perturbation $\theta_{var}(s)$  can be written as
\begin{equation}\label{varia2}
\delta^2\mathcal{V}=-B\int_{0}^{l}{\bigg[\theta_{var}''(s) +\gamma^2\,\theta_{var}(s)\,\cos{\big(\theta(s)-\beta\big)}\bigg]\,\theta_{var}(s)}\,\textup{d}s,
\end{equation}
so that an equilibrium configuration is stable when $\delta^2\mathcal{V}$ is positive definite ($\delta^2\mathcal{V}>0$ for every compatible perturbation $\theta_{var}(s)$), while
unstable when $\delta^2\mathcal{V}$ is indefinite or negative definite. Differently, in the case that $\delta^2\mathcal{V}$ is semi-positive definite,  higher order variations have to be considered
(as described at the end of this subsection)
to judge the stability.

The compatibility restricts the perturbation field $\theta_{var}(s)$ to have null values at the two ends, $\theta_{var}(0)=\theta_{var}(l)=0$,
and to satisfy the following isoperimetric constraints (derived from the first variation of the displacement boundary conditions
$x(l)=d$ and $y(l)=0$, eqns (\ref{BCkinematics})$_1$ and (\ref{BCkinematicdef}))
\begin{equation}\label{isoperim}
\int_{0}^{l}{\theta_{var}(s)\sin{\theta(s)}}\, \textup{d}s=0,\qquad
\int_{0}^{l}{\theta_{var}(s)\cos{\theta(s)}}\, \textup{d}s=0.
\end{equation}

Introducing the eigenfunction $\phi_n(s)$, subject to the  boundary conditions
\begin{equation}\label{BCphi}
\phi_n(0)=0,\quad \phi_n(l)=0,\quad \int_{0}^{l}{\phi_n(s)\sin{\theta(s)}}\, \textup{d}s=0,\quad \int_{0}^{l}{\phi_n(s)\cos{\theta(s)}}\, \textup{d}s=0,
\end{equation}
the positive definiteness of the functional $\delta^2\mathcal{V}$, eqn (\ref{varia2}), can be analyzed investigating the non-trivial solutions of the following Sturm-Liouville problem:
\begin{equation}\label{SLproblem}
\phi_n''(s)+\zeta_n\,w(s)\,\phi_n(s)=C_3 \sin{\theta(s)}+C_4 \cos{\theta(s)},
\end{equation}
where the weight function is defined as
\begin{equation}
w(s)=\gamma^2\cos{\big(\theta(s)-\beta\big)},
\end{equation}
and where the resultant inclination $\beta$, the normalized load $\gamma$ and the rotation field  $\theta(s)$ are known at this stage from the definition of equilibrium configuration,
while $C_3$ and $C_4$ are Lagrangian multipliers, and $\zeta_n$ is the eigenvalue related to the eigenfunction $\phi_n(s)$. The following property holds:
\begin{equation}\label{condfour}
\zeta_n \int_{0}^{l}{w(s)}\, \phi_n(s)\phi_m(s)\,\textup{d}s=\left\{
\begin{array}{lll}
\ds \zeta_n\int_{0}^{l}{w(s)\phi^2_n(s)}\, \textup{d}s=\int_{0}^{l}{\phi'_n(s)^2}\, \textup{d}s>0, &\textup{if}\,n=m\\[4mm]
\ds 0 &\textup{if}\,n\neq m,
\end{array}
\right.
\end{equation}
so that any compatible perturbation $\theta_{var}(s)$ can be expressed as the Fourier series expansion of the eigenfunctions $\phi_n(s)$
\begin{equation}
\theta_{var}(s)=\sum_{n=1}^\infty c_n \phi_n(s),
\end{equation}
where $c_n$ are the Fourier coefficients. Considering  the properties (\ref{condfour})  and the Fourier series expansion for the perturbation field $\theta_{var}(s)$,
the second variation of the total potential energy (\ref{varia2}) can be rewritten as
\begin{equation}\label{varia2fin}
\delta^2\mathcal{V}
=\sum_{n=1}^\infty \left(1-\frac{1}{\zeta_n}\right) c_n^2 \int_{0}^{l}{\phi'_n(s)^2}\,\textup{d}s,
\end{equation}
so that the equilibrium configuration is stable whenever $\zeta_n \notin \left[0,1\right]$ for every $n$,  unstable when $\zeta_n \in \left(0,1\right)$ for at least one value of $n$, and to be investigated through higher-order variations otherwise.  It follows that the condition of smallest eigenvalue
greater than one implies that the configuration is stable.
Following Kuznetsov and Levyakov \cite{lev}, the eigenfunction $\phi_n(s)$ can be represented as the linear combination of the functions $\varphi_n^{(j)}(s)$ ($j=1,...,4$)
\begin{equation}\label{solgen2}
\phi_n(s)=\sum_{j=1}^4 C_j \varphi_n^{(j)}(s),
\end{equation}
where the four functions are  defined respectively as the solutions of the following four second order differential problems
\begin{equation}\label{solveeq2}\begin{array}{lll}
&\left\{
 \begin{aligned}
&{\varphi_n^{(1)}}''(s)+\zeta_n\,w(s)\,\varphi_n^{(1)}(s)=0,\\
&\varphi_n^{(1)}(0)=1,\\
&{\varphi_n^{(1)}}'(0)=0,
 \end{aligned}
\right.
\qquad\qquad\qquad
&\left\{
 \begin{aligned}
&{\varphi_n^{(2)}}''(s)+\zeta_n\,w(s)\,\varphi_n^{(2)}(s)=0,\\
&\varphi_n^{(2)}(0)=0,\\
&{\varphi_n^{(2)}}'(0)=1,
 \end{aligned}
\right.
\\[10mm]
&\left\{
 \begin{aligned}
&{\varphi_n^{(3)}}''(s)+\zeta_n\,w(s)\,\varphi_n^{(3)}(s)=\sin{\theta(s)},\\
&\varphi_n^{(3)}(0)=0,\\
&{\varphi_n^{(3)}}'(0)=0,
 \end{aligned}
\right.
\qquad\qquad\qquad
&\left\{
 \begin{aligned}
&{\varphi_n^{(4)}}''+\zeta_n\,w(s)\,\varphi_n^{(4)}(s)=\cos{\theta(s)},\\
&\varphi_n^{(4)}(0)=0,\\
&{\varphi_n^{(4)}}'(0)=0.
 \end{aligned}
\right.
\end{array}
\end{equation}

Considering the boundary condition (\ref{BCphi})$_1$, and the boundary conditions defined for each
function $\varphi_n^{(j)}(s)$ ($j=1,...,4$)  in the differential problems (\ref{solveeq2}), it follows that $C_1=0$ so that the evaluation of the function $\varphi_n^{(1)}(s)$  can be disregarded.
Imposing the remaining boundary conditions (\ref{BCphi})$_2$, (\ref{BCphi})$_3$, and (\ref{BCphi})$_4$,
the homogenous linear problem $\textbf{A}(\zeta_n)\textbf{C}=\textbf{0}$ is obtained for the vector $\textbf{C}=\left\{C_2, C_3, C_4\right\}$, where
\begin{equation}
\textbf{A}(\zeta_n)=
\begin{bmatrix}
\varphi_n^{(2)}(l) & \varphi_n^{(3)}(l) & \varphi_n^{(4)}(l) \\[2mm]
\ds \int_{0}^{l}{\varphi_n^{(2)}(s)\cos{\theta(s)}}\,\textup{d}s & \ds \int_{0}^{l}{\varphi_n^{(3)}(s)\cos{\theta(s)}}\,\textup{d}s & \ds \int_{0}^{l}{\varphi_n^{(4)}(s)\cos{\theta(s)}}\,\textup{d}s \\[2mm]
\ds \int_{0}^{l}{\varphi_n^{(2)}(s)\sin{\theta(s)}}\,\textup{d}s & \ds \int_{0}^{l}{\varphi_n^{(3)}(s)\sin{\theta(s)}}\,\textup{d}s & \ds \int_{0}^{l}{\varphi_n^{(4)}(s)\sin{\theta(s)}}\,\textup{d}s
\end{bmatrix},
\end{equation}
so that the eigenvalues $\zeta_n$ can be finally evaluated from the condition of vanishing determinant, $\det{\textbf{A}(\zeta_n)}=0$.
From the operational point of view, with reference to every equilibrium rotation field $\theta(s)$,
the differential system (\ref{solveeq2}) can be numerically solved as a function of $\zeta_n\in[0,1]$ and
the configuration judged stable when $\det{\textbf{A}(\zeta_n)}\neq0$ for $\zeta_n\in[0,1]$, while unstable if this condition is not fulfilled. It is noted that in the limiting case when
$\det{\textbf{A}(\zeta_n)}$ vanishes only for $\zeta_n=1$, the related equilibrium configuration is  unstable if the third variation is non-null
for the related eigenfunction $\phi_{n}(s)$,
while if this variation vanishes the analysis should consider higher-order variations.
In particular, if the next even variation is positive  (negative) for $\theta_{var}(s)=\phi_{n}(s)$, the configuration is stable (unstable). Otherwise, if
the next even variation is null for $\theta_{var}(s)=\phi_{n}(s)$, the next odd variation should be considered and if not null the configuration is unstable.
 In general, the $k$-th variation of the total potential energy can be written for $k\geq3$ as
\beq
\delta^k\mathcal{V}=\left\{
\barr{lll}
\ds(-1)^{\frac{k+1}{2}}  \gamma^2 B\int_{0}^{l}{\left[\theta_{var}(s)\right]^k\,\sin{\big(\theta(s)-\beta\big)}}\,\textup{d}s,\qquad\qquad &k\,\, \mbox{odd},
\\[4mm]
\ds(-1)^{\frac{k}{2}}\gamma^2 B\int_{0}^{l}{\left[\theta_{var}(s)\right]^k\,\cos{\big(\theta(s)-\beta\big)}}\,\textup{d}s,\qquad\qquad &k\,\, \mbox{even}.
\earr
\right.
\eeq

\section[solutions]{Number of stable solutions, bifurcations and universal snap conditions}

For any triad of parameters $\left\{d,\theta_0,\theta_l\right\}$,
the nonlinear system (\ref{sistPOSnoflex}) can be solved to numerically evaluate
the existing pairs of parameters ($\beta$ and $\xi$ for $m=0$, or $\beta$ and $\eta$ for $m\neq0$) associated with the possible stable equilibrium configurations
for such boundary conditions.
Reference is made solely to configurations with $m\in[0,2]$ because those  with $m>2$ are numerically found to be always unstable (although a general analytical proof seems awkward, \cite{love}). The number of stable solutions has been observed to vary from 1 to 3 within the kinematical parameter space defined by
$\left\{d,\theta_0,\theta_l\right\}$.\footnote{It is remarked that  the end rotations  $\theta_0$ and $\theta_l$ have no restriction on their value, being unlimited their difference $\theta_l-\theta_0=\int_0^l \theta'(s) \textup{d}s$. If the angles $\theta_0$ and $\theta_l$ were referred to the end inclinations (instead of being referred to the end rotations, as in the present analysis), their respective sets would be limited due to  angular periodicity, for example to $\theta_0\in(-\pi,\pi)$ and
$\theta_l\in(-\pi,\pi)$, and  more than 3 stable configurations could be found as solution to the same end inclinations problem \cite{ardentov}.  
} 
Furthermore, being the developed model referred to the case of strips with loads applied only at the two ends,
configurations of self-intersecting elastica are found through the presented numerical evaluation
for a set of boundary conditions, while the respective self-contact configurations between different parts of the strip are excluded (see Section \ref{selfself} for further details).

\begin{figure}[!htb]
\begin{center}
\includegraphics[width=160mm]{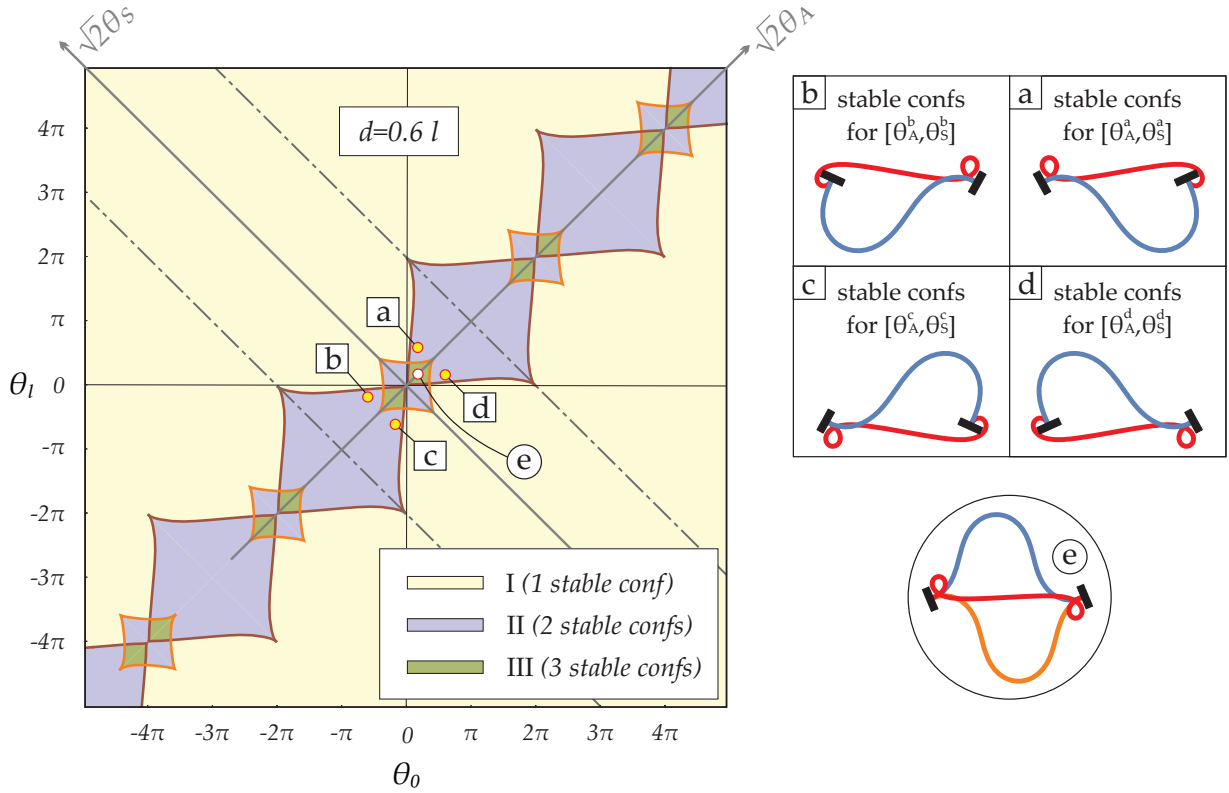}
\end{center}
\caption{\footnotesize (Left) Number of stable equilibrium configurations for a distance $d=0.6 l$ for varying $\theta_0$ and $\theta_l$ (or equivalently, $\theta_A$ and $\theta_S$).
Regions  with different colors and marked with I, II, and III identify triads $\left\{d,\theta_0,\theta_l\right\}$
for which  one, two, and three stable configurations exist, respectively.  (Right, upper part) Pairs of stable equilibrium configurations possible for the boundary conditions a, b, c, and d  highlighting the mirroring
properties when the modulus of the antisymmetric and the symmetric parts of the rotations is kept constant.
(Right, bottom part) The three stable equilibrium configurations possible for the boundary conditions e.
}
\label{fig_map06inclined}
\end{figure}

A typical map in the plane $\left[\theta_0,\,\theta_l\right]$ representing the number of stable solutions  is shown
in Fig. \ref{fig_map06inclined} (left) for a distance $d=0.6l$, where the existence of one, two, and three stable configurations for a specific triad $\left\{d,\theta_0,\theta_l\right\}$
is identified by the regions I, II, and III, respectively. The figure shows that the regions within this plane are characterized by a periodicity vector $[2\pi,2\pi]$ related
to the shifting in the solution of $2j\pi$ ($j\in \mathbb{Z}$) when the imposed rotations are modified by $2j\pi$ at both ends,
\beq\label{transfperiod}
\left\{
\barr{lll}
\theta^*(0)=\theta(0)+2 j \pi,\\[3mm]
\theta^*(l)=\theta(l)+2 j \pi,
\earr
\right.
\qquad
\Leftrightarrow
\qquad
\theta^*(s)=\theta(s)+2 j \pi, \qquad
\forall j \in \mathbb{Z},
\eeq

With reference to such periodicity property, it is instrumental to
introduce the angles $\theta_A$ and $\theta_S$, respectively
defined as the antisymmetric and symmetric parts of the imposed rotations,
\beq\label{thetasymeasym}
\theta_A=\frac{\theta_0+\theta_l}{2}, \qquad
\theta_S=\frac{\theta_l-\theta_0}{2},
\eeq
which are reported in Fig. \ref{fig_map06inclined} (left) through grey axes inclined at an angle $\pi/4$ with respect to the axes $\theta_0-\theta_l$.
Considering the shifting of the rotation field expressed by eqn (\ref{transfperiod}) it follows that
\beq\label{shifting}
\theta_A^*=\theta_A+2j\pi,
\qquad
\theta_S^*=\theta_S,\qquad
\forall j\in \mathbb{Z},
\eeq
highlighting the mentioned periodicity property of the equilibrium configurations which is given only
in the $\theta_A$ variable within the $\theta_A-\theta_S$  reference system.\footnote{
The periodicity for the solution in the rotation field for the elastica is similar to that observed in the kinematic description of the physical pendulum, which is insensitive to
an increase of an angle $2j\pi$ ($j\in \mathbb{Z}$).}

Another interesting property is now pointed out. A change in sign for each of the two parameters $\theta_A$ and $\theta_S$ is  related to
a specific \lq mirror' in the boundary conditions for the parameters $\theta_0$ and $\theta_l$, namely
\beq
\label{mirrorconds}
\begin{array}{lll}
&\mbox{Conf. $\mathrm{a}$:}\,\,
\left\{
\begin{array}{ccc}
\theta_A^{\mathrm{a}}=\overline{\theta}_A\\[3mm]
\theta_S^{\mathrm{a}}=\overline{\theta}_S
\end{array}
\right\}
\Leftrightarrow
\left\{
\begin{array}{ccc}
\theta_0^{\mathrm{a}}=\overline{\theta}_0\\[3mm]
\theta_l^{\mathrm{a}}=\overline{\theta}_l
\end{array}
\right\},
\qquad
&\mbox{Conf. $\mathrm{b}$:}\,\,
\left\{
\begin{array}{ccc}
\theta_A^{\mathrm{b}}=-\overline{\theta}_A\\[3mm]
\theta_S^{\mathrm{b}}=\overline{\theta}_S
\end{array}
\right\}
\Leftrightarrow
\left\{
\begin{array}{ccc}
\theta_0^{\mathrm{b}}=-\overline{\theta}_l\\[3mm]
\theta_l^{\mathrm{b}}=-\overline{\theta}_0
\end{array}
\right\},
\\[6mm]
&\mbox{Conf. $\mathrm{c}$:}\,\,
\left\{
\begin{array}{ccc}
\theta_A^{\mathrm{c}}=-\overline{\theta}_A\\[3mm]
\theta_S^{\mathrm{c}}=-\overline{\theta}_S
\end{array}
\right\}
\Leftrightarrow
\left\{
\begin{array}{ccc}
\theta_0^{\mathrm{c}}=-\overline{\theta}_0\\[3mm]
\theta_l^{\mathrm{c}}=-\overline{\theta}_l
\end{array}
\right\},
\qquad
&\mbox{Conf. $\mathrm{d}$:}\,\,
\left\{
\begin{array}{ccc}
\theta_A^{\mathrm{d}}=\overline{\theta}_A\\[3mm]
\theta_S^{\mathrm{d}}=-\overline{\theta}_S
\end{array}
\right\}
\Leftrightarrow
\left\{
\begin{array}{ccc}
\theta_0^{\mathrm{d}}=\overline{\theta}_l\\[3mm]
\theta_l^{\mathrm{d}}=\overline{\theta}_0
\end{array}
\right\}.
\end{array}
\eeq
It follows that the respective deformed configurations of the cases
$\mathrm{b}$, $\mathrm{c}$ and $\mathrm{d}$ can be obtained through the mirroring of the reference deformed
configurations of the case $\mathrm{a}$, Fig. \ref{fig_map06inclined} (right, upper part).
The mirroring properties can be summarized as:
\begin{itemize}
\item a change in sign for the parameter $\theta_A$
defines a configuration obtained as the mirroring with respect to the line orthogonal to that joining the two ends and passing at its center;
\item a change in sign for the parameter $\theta_S$
defines a configuration obtained as two mirrorings, one with respect to the line joining the two ends
and the other with respect to the orthogonal line at its mid point;
\item a change in sign  for both the parameters $\theta_A$ and $\theta_S$
defines a configuration obtained as the mirroring with respect to the line joining the two ends.
\end{itemize}

\begin{figure}[!htb]
\begin{center}
\includegraphics[width=140mm]{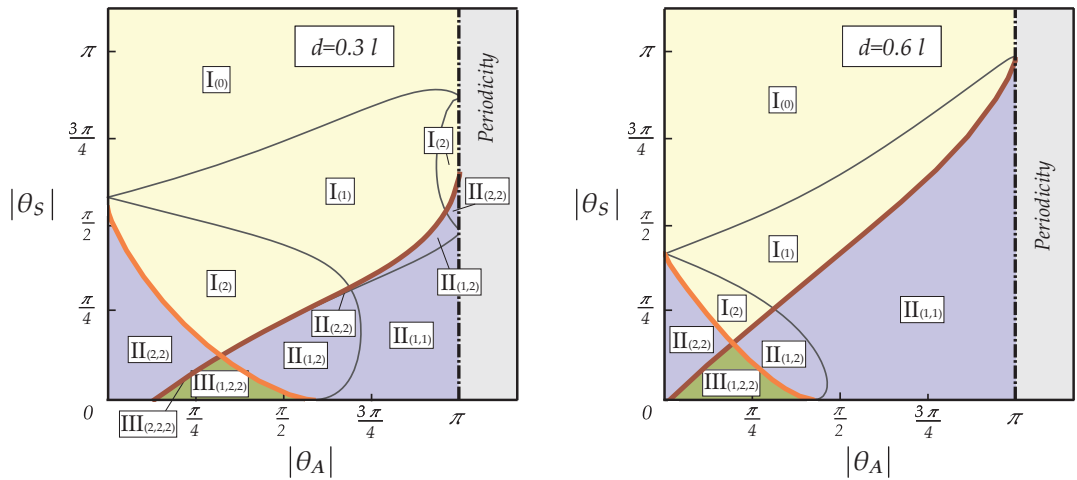}
\end{center}
\caption{\footnotesize Number of stable equilibrium configurations in the plane $|\theta_A|-|\theta_S|$, with the specification
of the inflection points number $m$ related to each stable solution listed in the subscript within parentheses, for the distances  $d=0.3l$ (left) and $d=0.6 l$ (right).
}
\label{fig_map06abs}
\end{figure}

These mirroring properties allow a simplified representation of the number of stable solutions within the plane $|\theta_A|-|\theta_S|$, so that only the first quadrant is drawn
and restricted to the condition $|\theta_A|\leq\pi$ because of the periodicity vector.
In this way, the map of  solutions number  reported in Fig. \ref{fig_map06inclined} (left) for a distance $d=0.6l$ is represented in Fig. \ref{fig_map06abs} (right), where
the inflection points number $m$ related to each stable solution is specified as listed in the subscript within parentheses.
As a further example, the representation of the number of stable solutions is also reported for a distance $d=0.3l$ in Fig. \ref{fig_map06abs} (left). Lines crossing and lines bounding the regions I, II, and III are reported in Fig. \ref{fig_map06abs}, in particular:
\begin{itemize}
\item the grey thin lines crossing the regions define the transition for which the number $m$ of inflection points  along the strip  changes, while the number of stable solutions is kept constant.
These lines can be defined imposing null curvature at one of the two ends (so that the strip has one hinged end while the other is a rotating clamp);
\item the thick lines bounding the regions define the transition for which the number of stable solutions changes, so that they represent the critical condition
of snap for one of the stable solutions. The lines are reported as thick orange and thick brown in order to identify the two possible different
snap-back conditions (described in the following).
\end{itemize}
Because of the simple reference (clamped-hinged) structure to which are related, the grey lines can be found from straightforward computations.
Indeed, for a given distance $d$, the relation $\theta_S=\theta_S(\theta_A)$ tracing the grey line can be computed from the nonlinear system (\ref{sistPOSnoflex}) by imposing
the condition of null moment at one of the two strip's ends, for example considering $\omega_0=(-1)^p \pi/2$ if the hinge is located at the coordinate $s=0$. Differently, drawing  the  orange and brown thick lines is a more
complex task because they are related to the issue of stability loss for one of the possible equilibrium configurations. Such analysis requires a further
representation for the solution domains which is now introduced.

\subsection{Equilibrium paths for a fixed distance $d$}

\begin{figure}[!h]
\begin{center}
\includegraphics[width=160mm]{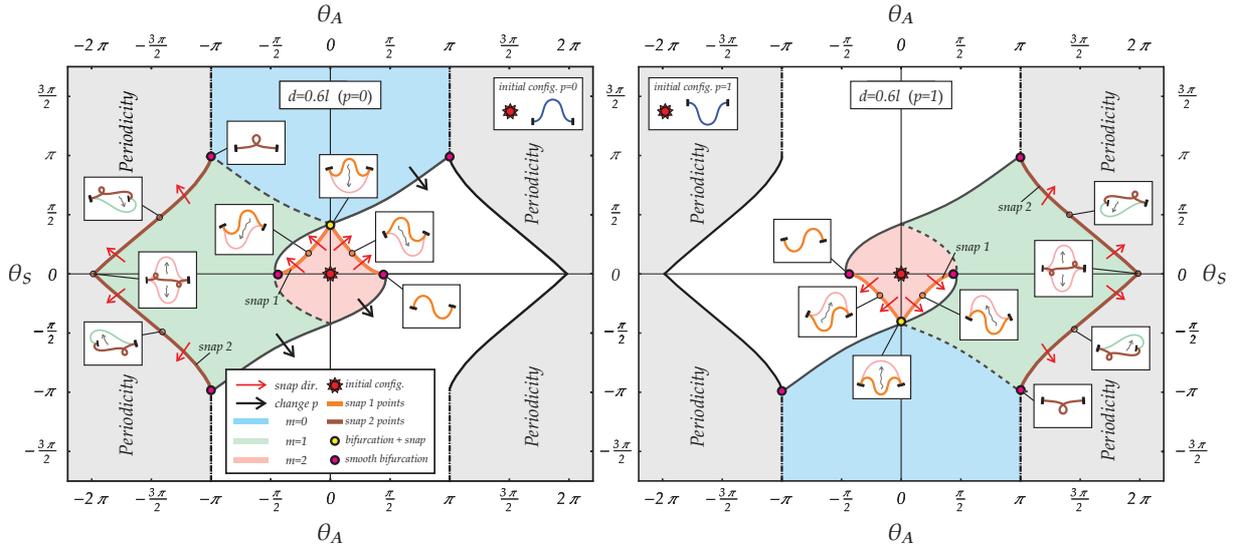}
\end{center}
\caption{\footnotesize Domains of unique stable solutions within the plane $\theta_A-\theta_S$ for $p=0$ (left) and $p=1$ (right)
for evolutions with a fixed distance $d=0.6 l$ and starting from the boundary conditions $\theta_A=\theta_S=0$.
Regions with the same color identify
stable equilibrium configurations having the same number $m$ of inflection points  along the strip  (note the mirroring property of the domains for varying $p$).
See the main text for the definition of the domain and line colors.
}
\label{fig_incremdxp06}
\end{figure}

In the attempt to simplify the visualization of the solution domains for a fixed distance $d$, reference is made to the quasi-static evolutions
of the rotations $\theta_A$ and $\theta_S$ starting from a stable configuration for null angles at both clamps, $\theta_0=\theta_l=0$ so that $\theta_A=\theta_S=0$ (namely, a shortening of the distance between the two clamps is imposed starting from the straight configuration).
Only two stable solutions can be obtained for these \lq initial' boundary conditions, both characterized by
the same number of inflection points ($m=2$) but differing in the value of $p$, and representative
of the two possible buckled configurations for a strip clamped at both ends and subject to a shortening $l-d$.
Considering this, the solution maps of Fig. \ref{fig_map06inclined} and of Fig. \ref{fig_map06abs} (right) for a fixed distance $d=0.6 l$ can be split in two separate representations, each of these
 restricted to a specific value of $p$ (identifying the positiveness, $p=0$, or negativeness, $p=1$, of the curvature at the coordinate $s=0$) as shown in Fig. \ref{fig_incremdxp06}
 and therefore related to one of the two initial configurations possible for $\theta_A=\theta_S=0$.
In this way, the generic evolution for the boundary conditions can be represented with a continuous curve in the plane $\theta_A-\theta_S$ starting from its origin. Despite the non-linearity of the problem, each point along this curve is related to a (when existing) unique deformed configuration whenever snap conditions or bifurcation points are not encountered along the considered evolution. Consequently, such a representation provides a fundamental tool in the definition of the solution domains and of the snap-back conditions for
every possible evolution of the imposed rotations at a fixed distance $d$, showing the critical and bifurcation conditions depending on the value of $p$.

The moment-rotation response curve  associated with each evolution of the boundary conditions reveals the presence of  snap-back instabilities (related to the annihilation of the second variation,
 $\delta^2\mathcal{V}=0$) when a point with vertical tangent in the response curve is reached and no smooth stable\footnote{In general, the equilibrium path smoothly continues after the snap point with an opposite change of the  rotation value. However, this path is not considered here due to its unstable nature.} evolutions of the deformed configuration can be obtained for a further monotonic variation in the rotation.
The set of boundary conditions $\left\{d, \theta_A, \theta_S\right\}$ corresponding to
the vertical tangency in the moment-rotation response are identified by means of a standard bisection algorithm applied to the analyzed equilibrium path.

The above described analysis provides the domains and lines as reported in Fig. \ref{fig_incremdxp06}, where also specific equilibrium configurations are displayed for some critical pairs of boundary conditions. The solution domains are reported with different colors, identifying different properties as follows:
\begin{itemize}
\item
    \textit{blue/green/red regions} -- the unique stable equilibrium configuration corresponding to the considered $p$ has
    zero, one, and two inflection points along the strip  for the
    blue, green, and red regions, respectively;
\item
    \textit{white regions} -- no stable equilibrium configuration is possible for that specific value of $p$. However, for the mirroring property, the stable equilibrium configuration
    exists for the other value of $p$;
\item
    \textit{grey regions} -- the equilibrium configuration related to a pair $\overline{\theta}_A, \overline{\theta}_S$ belonging to this region can be represented by that related
    to the pair within the colored or white regions (namely, outside the grey regions) considering the shifting of the solution as highlighted by eqn (\ref{shifting}).
    The  corresponding dual values can be evaluated from $\overline{\theta}_A$ and $\overline{\theta}_S$ as
    \beq\label{shiftingrule}
    \theta_A=\overline{\theta}_A-2\pi\left\lfloor \frac{1}{2}+\frac{\overline{\theta}_A}{2\pi}\right\rfloor,\qquad \theta_S=\overline{\theta}_S,
    \eeq
    where the symbol $\lfloor \cdot \rfloor$ represents the floor function, which evaluates the
    greatest integer that is less than or equal to the relevant argument. Note that, from eqn (\ref{shiftingrule}) follows that $\theta_A\in [-\pi,\pi]$ for every $\overline{\theta}_A$.
\end{itemize}

Similarly, the lines separating the different domains are drawn with various styles, representing
the type of transition in the equilibrium configuration occurring when the evolution of the boundary conditions passes from one region to another.

With reference to any continuous evolution in the set of applied rotations $\left\{\theta_A(\tau),\,\theta_S(\tau)\right\}$ from an initial ($\tau=\tau_i$)
to a final instant ($\tau=\tau_f$), where $\tau$ is a time-like parameter,
the following cases are possible:
\begin{itemize}
\item
    if a \textit{dashed grey line} is crossed,  the equilibrium configuration smoothly varies
    changing the number $m$ of the inflection points, while the value of $p$ is kept constant;
\item
    if a \emph{continuous grey line} is crossed,
    the equilibrium configuration smoothly varies changing both the number $m$ of the inflection points and the
    value of $p$, in particular $p(\tau_f)=1-p(\tau_i)$.  Therefore,
    the final configuration is related to a solution map dual to that of the initial value of $p(\tau_i)$. As it can be noted in Fig. \ref{fig_incremdxp06},
    the continuous grey lines are present only at the borders between one of the colored regions with a white region;
\item
    if a \emph{dashdotted grey line} is crossed,
    the equilibrium configuration smoothly varies keeping fixed both the number $m$ of the inflection points and the
    value of $p$. As it can be noted in Fig. \ref{fig_incremdxp06},
    these lines appears only at $\theta_S=\pm\pi$ and are connected  to the shifting  property of the solution, eqn (\ref{shifting}), so
    that, considering eqn (\ref{shiftingrule}), the final equilibrium configuration
    has to be referred to the values
     $\left\{\theta_A(\tau_f)-2\pi\lfloor1/2+\theta_A(\tau_f)/(2\pi)\rfloor\right.,$ $\left.\theta_S(\tau_f)\right\}$;
\item
    if a \emph{thick orange line} is crossed, a critical configuration of snap-back type 1 instability is encountered
    so that a small variation in the boundary conditions realizes a large variation in the equilibrium configuration.
    Therefore, the final equilibrium configuration is achieved by means of a dynamic motion from the initial one. The initial and final configurations are described
    by a different value of $p$, namely, $p(\tau_f)=1-p(\tau_i)$. From the present analysis it is also observed that before and after the
    snap-back type 1 mechanism there are always two inflection points,  $m(\tau_f)=m(\tau_i)=2$;
\item
    if a \emph{thick brown  line} is crossed,
    the critical condition of snap-back type 2 instability  is encountered. However, differently from  crossing the thick orange line, in this case the boundary conditions of the final configuration
    $\left\{\theta_A(\tau_f)/(2\pi),\theta_S(\tau_f)\right\}$ lie within the grey region, so
    that the interpretation of the considered solution map requires  a further effort. Indeed, from  the solution shifting principle,
    the final configuration should be referred to the values $\left\{\theta_A(\tau_f)-2\pi\lfloor1/2+\theta_A(\tau_f)/(2\pi)\rfloor,\theta_S(\tau_f)\right\}$.
    If these values correspond
    to a white region for the solution map with $p(\tau_i)$, the final configuration is the one associated with the same values but related to the dual map for which $p(\tau_f)=1-p(\tau_i)$;
\item
    if no line is crossed, the initial and final configurations have the same values for $p$  and $m$, and therefore are represented within the same solution map  corresponding to $p(\tau_i)=p(\tau_f)$.
\end{itemize}

It is worth to remark that  crossing a colored thick line does not always provide a snap-back instability. Indeed, this phenomenon is strictly related to the equilibrium configuration
taken by the structural system before crossing this condition.
More specifically, referring to the $\theta_A-\theta_S$ plane and a fixed distance $d$, the snap mechanism is realized whenever the scalar product between the incremental vector connecting the initial to the final boundary conditions and 
the normal (defined as the derivative of the tangent) to the snap-back curve is non-negative.

Moreover, the snap-back curves display the typical shape of catastrophic cusps \cite{bazant}, revealing how the magnitude of the so-called control
parameter at a critical point decreases with respect to the perfect case (maximum critical
rotation at the cusp) due to the presence of increasing imperfections. In particular, the role of such control parameter is played by the
symmetric part of the rotation $\theta_S$ for snap-back curves of type 1 and by the antisymmetric part $\theta_A$ for snap-back curves of type 2, while the role of the imperfection
is respectively played by $\theta_A$ and $\theta_S$.

It is finally noted that all the regions and lines reported in Fig. \ref{fig_incremdxp06} satisfy the
mirroring properties shown in eqns (\ref{mirrorconds}) and highlighted in Fig. \ref{fig_map06inclined} (right). Therefore,
when $p$ is switched from 0 to 1, the domains of the case $p=1$ (Fig. \ref{fig_incremdxp06}, right) can be
obtained from those of the case $p=0$  (Fig. \ref{fig_incremdxp06}, left) through a mirroring
with respect to the axis $\theta_S$ for the configurations with $m=1$ and through a double mirroring
(with respect to both axes, $\theta_A$ and $\theta_S$) for those with $m=0$ and $m=2$.

\begin{figure}[!htb]
\begin{center}
\includegraphics[width=160mm]{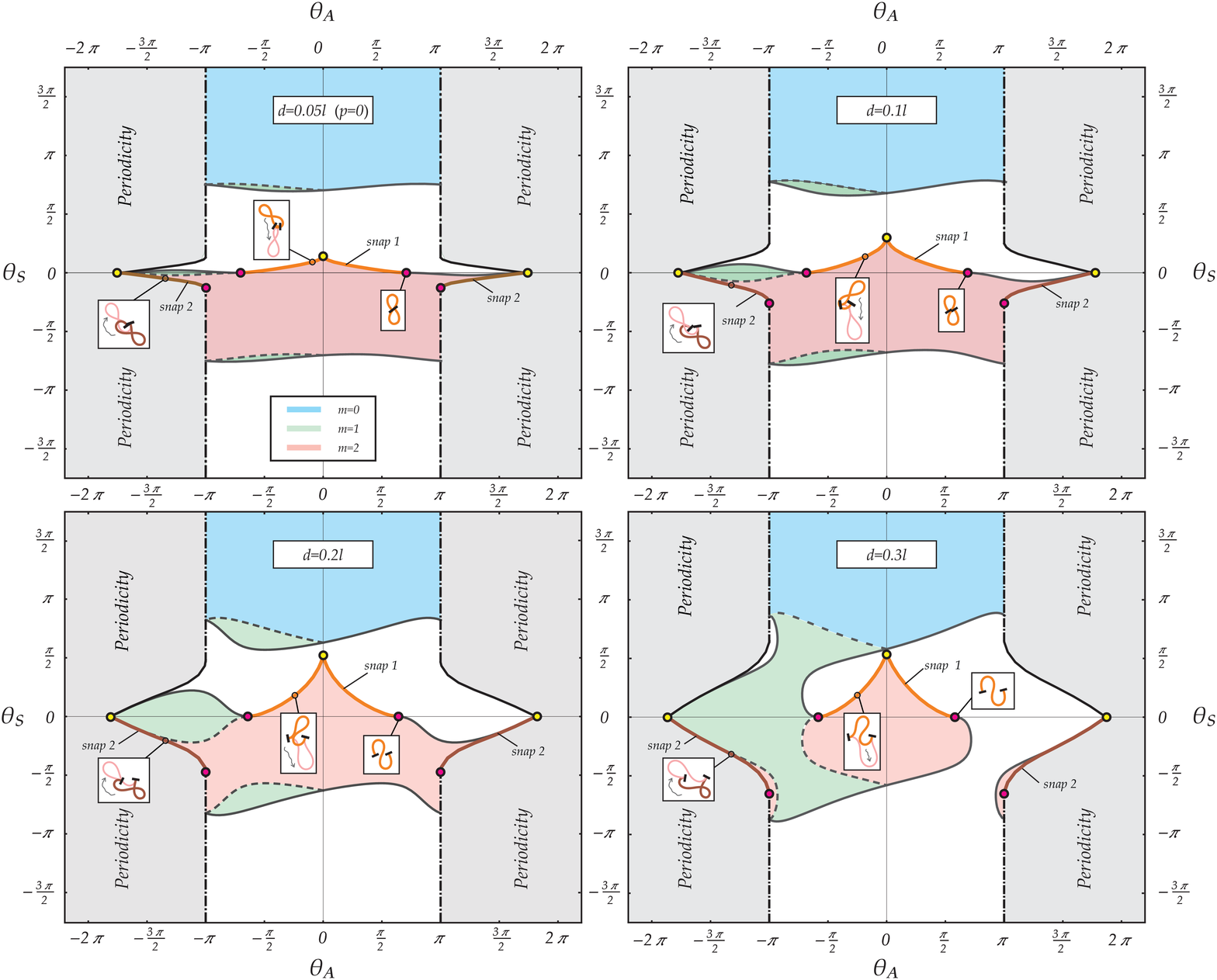}
\end{center}
\caption{\footnotesize Domains of unique stable solutions within the plane $\theta_A-\theta_S$
for evolutions starting from the boundary conditions $\theta_A=\theta_S=0$ and related to $p=0$ (the domains related to $p=1$ can be obtained through
the respective mirroring properties highlighted in the text and visible in Fig. \ref {fig_incremdxp06}).
The domains are reported for different values of fixed distance between the two ends, $d=\left\{0.05, 0.1, 0.2, 0.3\right\} l$.
See the main text for the definition of the domain and line colors.
}
\label{fig_collectiond1}
\end{figure}

\begin{figure}[!htb]
\begin{center}
\includegraphics[width=160mm]{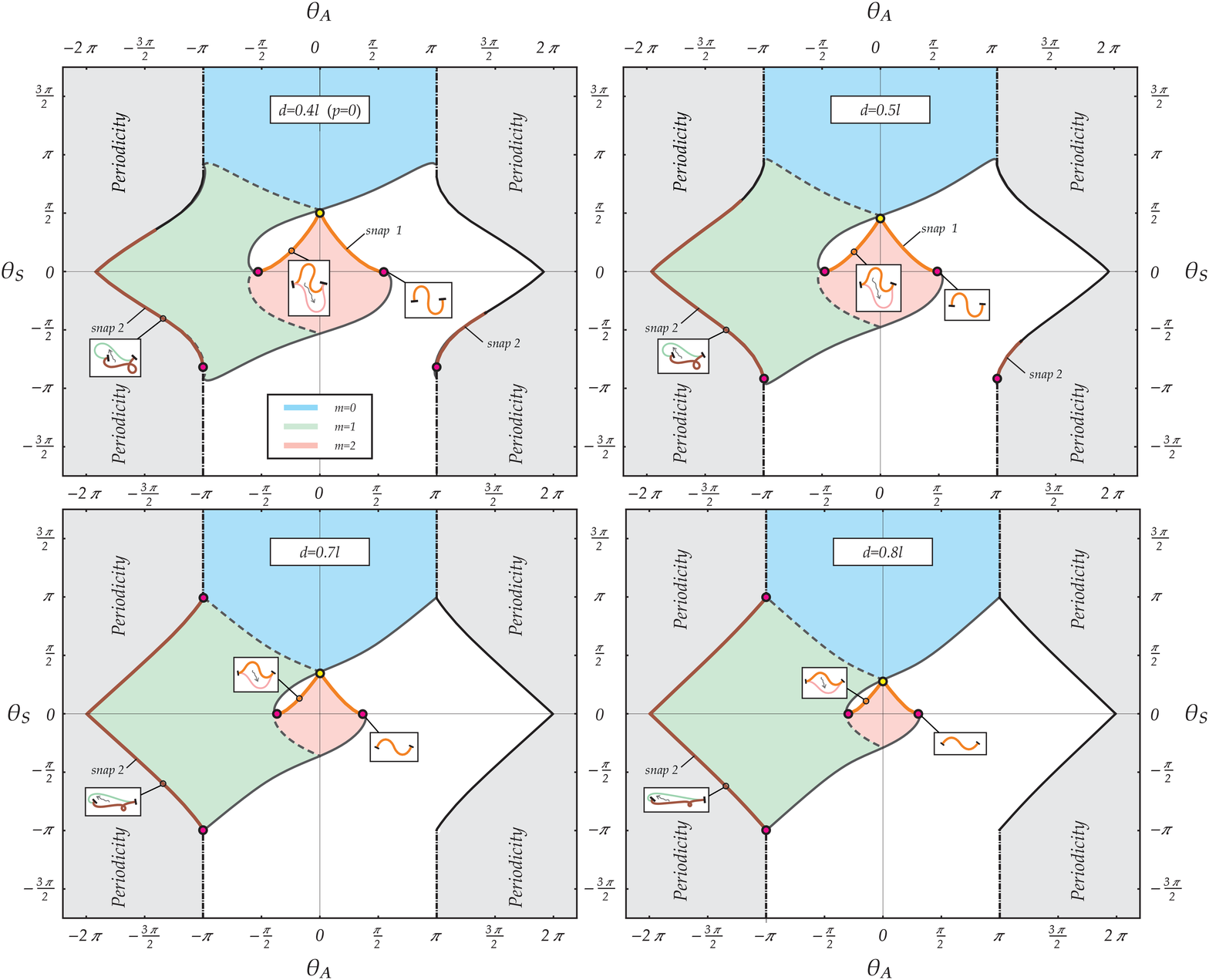}
\end{center}
\caption{\footnotesize As for Fig. \ref{fig_collectiond1}, but for higher values of fixed distance between the two ends, $d=\left\{0.4,
0.5, 0.7, 0.8\right\} l$. See the caption of Fig. \ref {fig_incremdxp06}
and the main text for the definition of the domain colors and the other lines. Being $d> 0.369 l$, the yellow points (associated with bifurcations)
are absent at the cusps of the snap-back curves type 2 (see Sect. \ref{bifurcs}).
}
\label{fig_collectiond2}
\end{figure}

Similarly to Fig. \ref{fig_incremdxp06}, the domains are reported in Figs. \ref{fig_collectiond1} and \ref{fig_collectiond2}
for different values of the end's distance, respectively for $d=\left\{0.05, 0.1, 0.2, 0.3 \right\}l$ and $d=\left\{0.4, 0.5, 0.7, 0.8 \right\}l$.
The maps are reported only for $p=0$, since those are related to the case $p=1$ through mirroring.

A last comment is made about the two limit cases of maximum and minimum  distance between the ends, respectively $d\simeq l$ and $d=0$:
\begin{itemize}
\item In the former case ($d\simeq l$), although the analysis should be improved by considering stretching energy,
the present model (based on the inextensibility assumption) shows that the curve of snap-back type 1 reduces to the point with coordinates $\theta_A=\theta_S=0$
 while the curve of snap-back type 2 reduces to the four line segments defined by $|\theta_S|=2\pi-|\theta_A|$ with $|\theta_A|\in[\pi,2\pi]$;
\item Differently, in the case when the two ends have the same position  ($d=0$),  the mechanical system shows the independence of the parameter $\theta_A$, because in the case of null distance
the angle $\theta_A$ merely expresses a rigid rotation of the entire structure and the elastic energy stored within the structure is only a function of the angle $\theta_S$.
In this case, the curve of snap-back type 1 becomes the segment lines given by $\theta_S=0$ and $|\theta_A|\lesssim 0.726\pi$ while the curve of snap-back type 2 becomes the segment lines given by  $\theta_S=0$
and $|\theta_A|\in[1,1.726]\pi$.
It is also worth to mention that in the very special case of $\theta_S=0$, an infinite set of stable and equivalent (namely, corresponding to the same elastic energy) solutions exists
for the strip with fixed end rotations, provided by a \lq 8-shaped' configuration \cite{lev8,sachkov}.

\end{itemize}

\subsection{Equilibrium paths for a variable distance $d$}

\begin{figure}[!htb]
\begin{center}
\includegraphics[width=140mm]{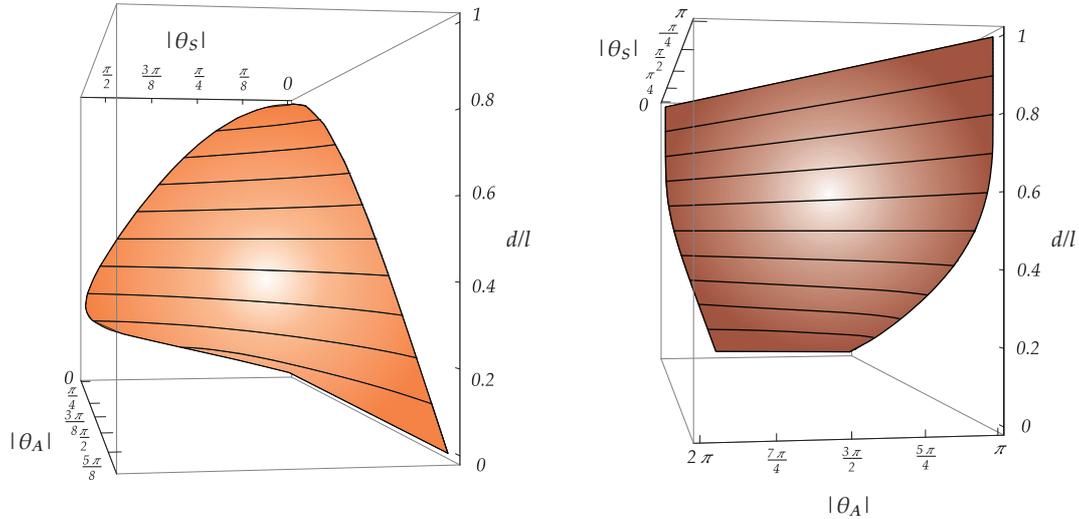}
\end{center}
\caption{\footnotesize
Surfaces defining the critical boundary conditions in the $\left\{|\theta_A|,|\theta_S|, d/l\right\}$ space for which snap-back type 1 (left) and type 2 (right) occurs.
The surfaces are  built from the related
snap-back curves, respectively orange and brown, obtained for different values of $d$ and reported in Figs. \ref{fig_incremdxp06}, \ref{fig_collectiond1}, and \ref{fig_collectiond2}.
} \label{snap3d}
\end{figure}

The families of snap curves reported for fixed values of $d$ in Figs.  \ref{fig_incremdxp06}, \ref{fig_collectiond1} and  \ref{fig_collectiond2}  suggest the existence of snap surfaces within the space $d/l-\theta_A-\theta_S$. 
These snap surfaces can be disclosed by interpolating the snap curves within specific planes. The snap curves evaluated for twenty planes, taken for computational convenience at constant $\theta_A$ and  $\theta_S$  for snap types 1 and 2, respectively, have been exploited to generate the snap-back surfaces by means of the function $\texttt{Interpolation}$ in \textit{Mathematica} (v.10). Due to mirroring properties, the snap surfaces are  entirely described through their representations within one octant of the space $d/l-\theta_A-\theta_S$,  displayed in  Fig. \ref{snap3d} for snap-back type 1 (left) and type 2 (right) surfaces. 
Such snap surfaces represent a fundamental tool in the investigation about  snap mechanisms and the definition of the correspondent critical boundary conditions during a
 quasi-static evolution of strips with controlled ends.
 With the purpose to facilitate the use of this tool, the numerically evaluated dataset of  about 2000 critical conditions for snap type 1 is made available as Supplementary material.
 Before exploiting the concept of universal snap surfaces in some applicative examples presented in Sect. \ref{lastsect},
it is worth to discuss the possibility of bifurcations during a loading process, the amount of elastic energy release, and the conditions of self-intersecting elastica.

\subsection{Bifurcations at snap}\label{bifurcs}

The boundary conditions corresponding to  (almost all) the intersections of the orange and brown snap-back curves
with the lines $\theta_A=\left\{-\pi,0,\pi\right\}$ and  with the line $\theta_S=0$ (corresponding to the cusps and end points of the snap curves)
are marked with colored spots in Figs. \ref{fig_incremdxp06}, \ref{fig_collectiond1}, and \ref{fig_collectiond2}. For the boundary conditions corresponding to these points,
the semi-positive definiteness of the second variation $\delta^2 \mathcal{V}$
 is complemented by the annihilation of the third variation $\delta^3 \mathcal{V}$ for the related eigenfunction
   (while {$\delta^3 \mathcal{V}$ is different from zero for the other
 points along the snap curve)
  and a positive value is found for the corresponding fourth variation $\delta^4 \mathcal{V}$.
  These special points (some of them also observed in \cite{manning}) are marked as red and yellow spots, 
respectively corresponding to pitchfork (with no snapping) and unstable-symmetric bifurcation points (with snapping), and defined as follows.

\paragraph{Red spots.} These points are located at the intersections of the orange thick curve with the condition $\theta_S=0$ and
 of the brown thick curve with the condition $\theta_A=\pm\pi$. Snap never occurs when  the snap curve is crossed through these points; however
when the snap curve is crossed from outside to inside, a bifurcation may be encountered. More specifically, a pitchfork bifurcation is found
 when crossing the orange (brown) curve from outside to inside at
 the red point with a null increment in the angle $\theta_S$ ($\theta_A$), namely a purely antisymmetric (symmetric) variation in the boundary
 conditions at the critical point. These bifurcative behaviours are displayed in Fig. \ref{bifurcation1} and Fig. \ref{bifurcation2}, where
 specific loading processes are considered for a fixed distance $d=0.4 l$ and crossing the \emph{red spots} located at the edges of the snap-back type 1 and type 2 curves.
 \begin{figure}[!htb]
\begin{center}
\includegraphics[width=180mm]{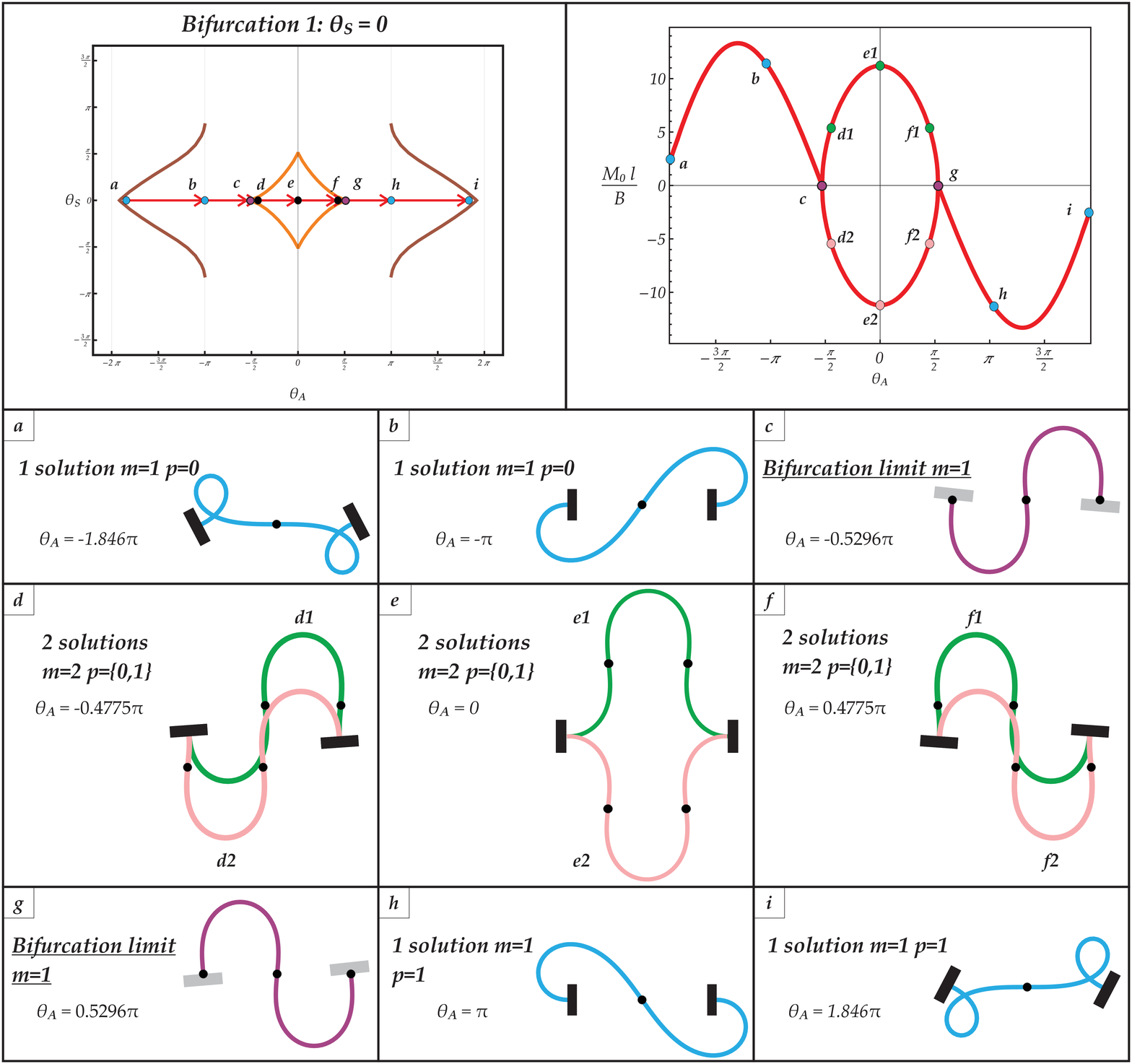}
\end{center}
\caption{
 \footnotesize Evolution of the left clamp moment $M_0$ for increasing $\theta_A$ (upper line, right column),
showing the existence of pitchfork bifurcation points at $c$ and $g$ along the antisymmetric loading path $\theta_S=0$ at fixed distance $d=0.4l$ (upper line, left column).
The evolutions of the stable equilibrium configurations along the considered path are shown in the second, third and fourth line.
}
\label{bifurcation1}
\end{figure}

\begin{figure}[!htb]
\begin{center}
\includegraphics[width=180mm]{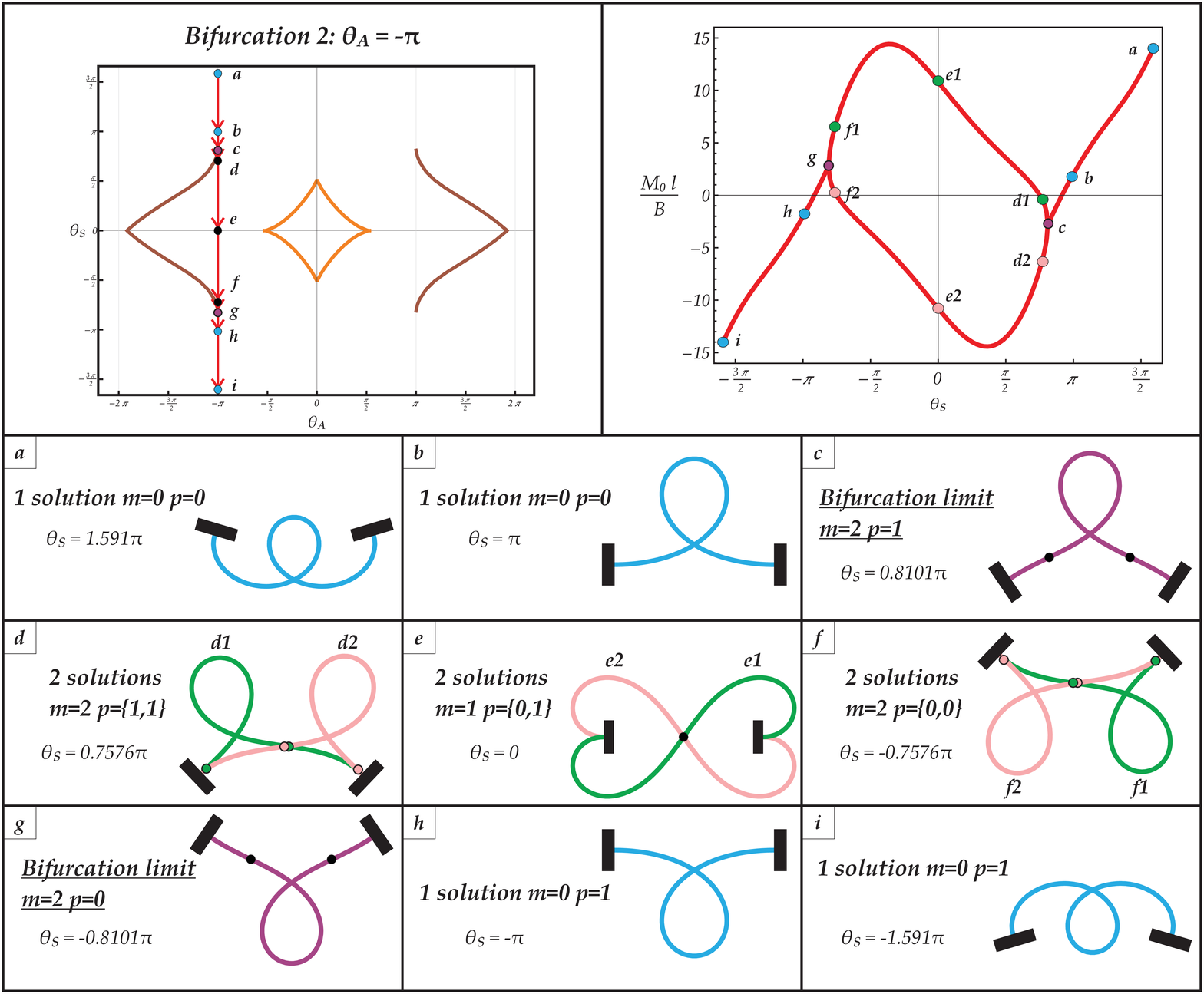}
\end{center}
\caption{\footnotesize As for Fig. \ref{bifurcation1}, but for varying $\theta_S$  with $\theta_A=-\pi$ and $d=0.4l$.
As the configuration $c$ is reached, a pitchfork bifurcation occurs at the limit configuration (purple, $m=2$) followed by symmetry-breaking in both branches, as shown by the
two stable configurations (green and pink) at stages $d$, $e$, and $f$.
}
\label{bifurcation2}
\end{figure}
  The antisymmetric case, $\theta_S=0$, at increasing $\theta_A$ is considered in Fig. \ref{bifurcation1},
where in the first line the loading path is reported on the left and the moment $M_0$ at the left clamp  as a function of the angle parameter $\theta_A$ on the right.
The evolution of the deformed shape is displayed in the second, third, and fourth lines and is reported for the six specific stages
 of the boundary conditions, highlighted both in the solution maps and moment-angle response through the alphabetic letters  $a$, $b$,  $c$, $d$,  $e$,  $f$, $g$, $h$, and $i$.
During the evolution, the bifurcation occurs when the stage $c$ is attained, namely, when the snap curve is crossed from outside to inside, and corresponds to the
condition of null moment at both clamps.
Just after the stage $c$ is passed, the structure may equally evolve through two different loading paths, namely, the structure may equally
reach configuration $d1$ or $d2$. Once that one of these two branches is undertaken, the evolution continues on that specific branch. However, at increasing the rotation parameter,
both branches finally join together when the stage $g$ is reached and, after this stage (from $g$ to $i$), the structure follows the only possible evolution.
It is also important to highlight that the configurations $c$ and $g$ are stable solutions having $m=1$ internal inflection point and null curvature at both ends (similar configurations are also present for all the bifurcation conditions, red spots, on snap-back curve type 1 for every distance $d$,
and for some bifurcation conditions, yellow spots, on snap-back curve type 2 as discussed below).

A similar behaviour is also displayed in  Fig. \ref{bifurcation2} with reference to  $\theta_A=-\pi$ and decreasing  $\theta_S$. Differently from Fig. \ref{bifurcation1}, here
the bifurcation occurs for a symmetric configuration and is associated with a non-null moment value at both clamps.
The pitchfork bifurcation at the point $c$ reveals a symmetry-breaking behaviour. In fact, the initial symmetric configuration becomes unstable after the bifurcation point is reached,
and the system may only follow two other stable paths described by the mirrored configurations. A similar behaviour has also been detected in the case of a ring pinched by two radial loads
\cite{lev}, while  the existence of a central unstable and symmetric solution has also been reported by \cite{bigonidef} and \cite{lev},
where the symmetric configuration for the double-clamped rod with null rotations at its ends
is proven to snap towards the S-shaped configurations $e1$ or $e2$.

\paragraph{Yellow spots.} These points are located at the intersections of the orange thick curve with the condition $\theta_A=0$ and, only for $d\leq 0.369\,l$,
of the brown thick curve with the condition $\theta_S=0$.\footnote{It is remarked that the yellow points  given by the intersection of the brown snap-back curve with the axis $\theta_S=0$
exist only for $d\leq 0.369\,l$ and correspond to stable configurations with $m=1$ inflection point and null curvature at both ends. This behaviour is not observed for $d> 0.369\,l$, where the third variation $\delta^3 \mathcal{V}$ is not null
and snap mechanism occurs without any bifurcation as soon as the snap curve is crossed at any other point. It follows that
no yellow point appears in Fig. \ref{fig_collectiond2} for snap-back curves of type 2, being the considered distances $d> 0.369\,l$.}
Snap occurs when the snap curve is crossed through these points from inside to outside
and a bifurcation may be even encountered. More specifically, an unstable-symmetric bifurcation is found at the snap
 when crossing the orange (brown) curve from inside to outside at
 the yellow point with a null increment in the angle $\theta_A$ ($\theta_S$), namely a purely symmetric (antisymmetric) variation in the boundary
 conditions at the critical point. The moment rotation response at increasing value of $\theta_S$ crossing the snap type 1 curve
is displayed in Fig. \ref{fig_mom3D} for a fixed distance $d=0.4 l$ and at very small fixed values of $\theta_A=\{0,10^{-4},10^{-3},10^{-2}\}$.
An example of unstable-symmetric behaviour can be envisaged in the purely symmetric case, $\theta_A=0$,
where the symmetric response (for which the moments at the two clamps have the same value, $M_0=M_l$) intersects the two other unstable paths. Differently, if a small constant value is assumed
for the antisymmetric part of rotation $\theta_A$, the moment rotation response reaches a critical condition of snap-back, for which the tangent of the response curve is vertical.

\begin{figure}[!htb]
\begin{center}
\includegraphics[width=160mm]{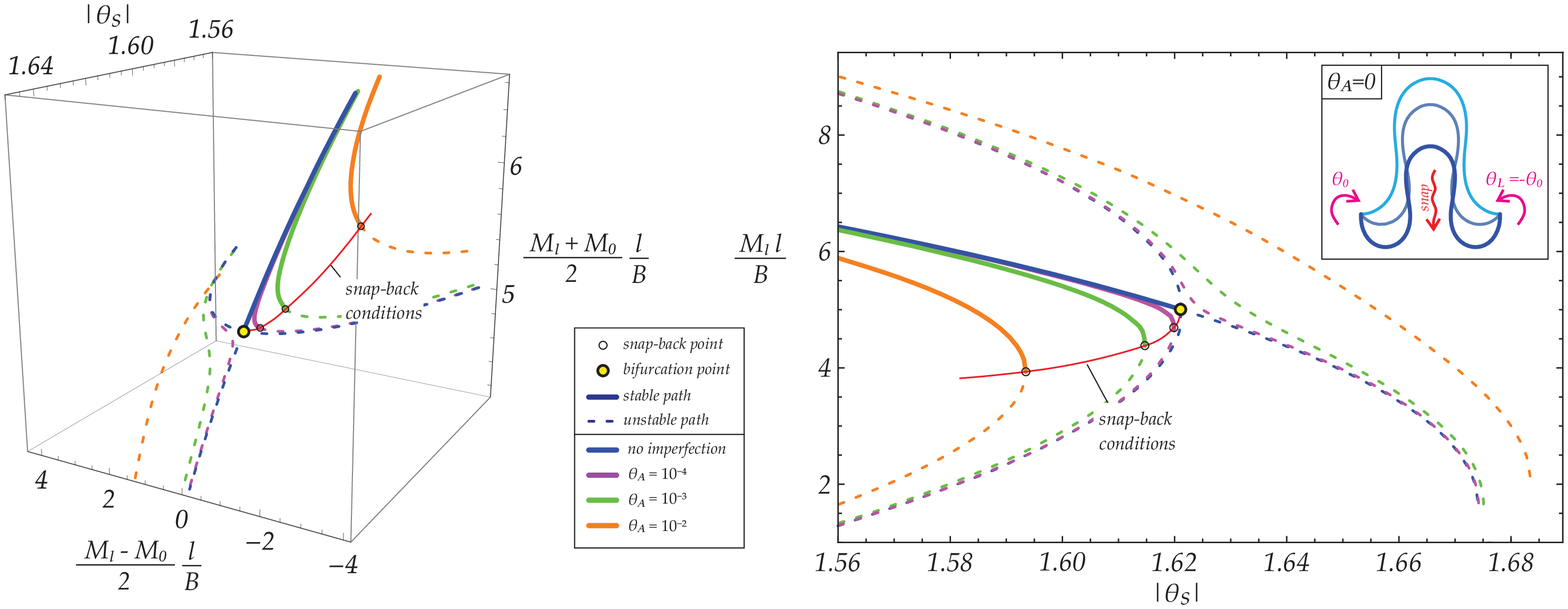}
\end{center}
\caption{
\footnotesize 
Moment-rotation responses at increasing modulus of $\theta_S$ (crossing the snap type 1 curve) for a fixed distance $d=0.4 l$ and $\theta_A=\left\{0,10^{-4},10^{-3},10^{-2}\right\}$.
The presence of an unstable-symmetric bifurcation point (yellow spot) is reported for symmetric loading condition, $\theta_A=0$.
(Left) Responses in terms of normalized symmetric and antisymmetric parts of the moment at the clamps, $(M_0+M_l)/2$ and $(M_0-M_l)/2$. (Right) Responses in terms of the normalized moment at the right clamp, $M_l$, and (inset) three deformed configurations before snapping and corresponding to $\theta_S=\{0, 0.9, 1.621\}$ and to $\theta_A=0$.
}
\label{fig_mom3D}
\end{figure}

\subsection{Self-intersecting elastica}\label{selfself}

The  snap surfaces reported in Fig. \ref{snap3d} have been obtained assuming that the strip is only loaded  at its ends. It follows that the obtained critical conditions
hold whenever the development of self-contact points  along the strip  is excluded.\footnote{Analysis of elasticae with self-contact points requires the resolution
of two or more elasticae subject only to end loadings \cite{plaut2004}.} This circumstance is trivially realized when the deformed
configuration is not self-intersecting, but also when the self-intersection is made possible by the out-of plane geometry.
The latter case is realized with strips shaped along  the out-of-plane direction in such a way that during the planar intersection two external halves
of the strip contain a  central strip, namely a Y-shaped strip with specific out-of-plane variations  \cite{bigoni, lev}.

In order to detect when self-intersection does occur before snapping and, equivalently, when it does not,
it is of practical interest to define the boundary conditions for which a contact point is first realized. This information can be consequently used to
 determine which portions of snap surfaces are attained only after
an evolution involving the self-intersection. The boundary conditions of first self-contact can be found imposing that for one and only one pair of curvilinear coordinates $s^{(1)}$ and $s^{(2)}$ have the
same position,
\beq
x\left(s^{(1)}\right)=x\left(s^{(2)}\right), \qquad
y\left(s^{(1)}\right)=y\left(s^{(2)}\right),\qquad \left\{s^{(1)}, s^{(2)} \right\}\in [0,l].
\eeq
Evaluating the conditions of first self-contact, it is observed that:
\begin{itemize}
\item the conditions of snap-back type 1 (orange surface, Fig. \ref{snap3d} left) are always reached without developing a self-intersecting elastica if $d\gtrsim l/4$.
For $d\lesssim l/4$, the snap-back most likely occurs after developing a self-intersecting configuration,  however the exact limit distance for which the self-intersection
is realized depends on the values of $\theta_A$ and $\theta_S$, Fig. \ref{fig_selfcomp};
\item the conditions of snap-back type 2 (brown surface, Fig. \ref{snap3d} right) are always reached after developing a self-intersecting elastica.
\end{itemize}

\begin{figure}[!htb]
\begin{center}
\includegraphics[width=160mm]{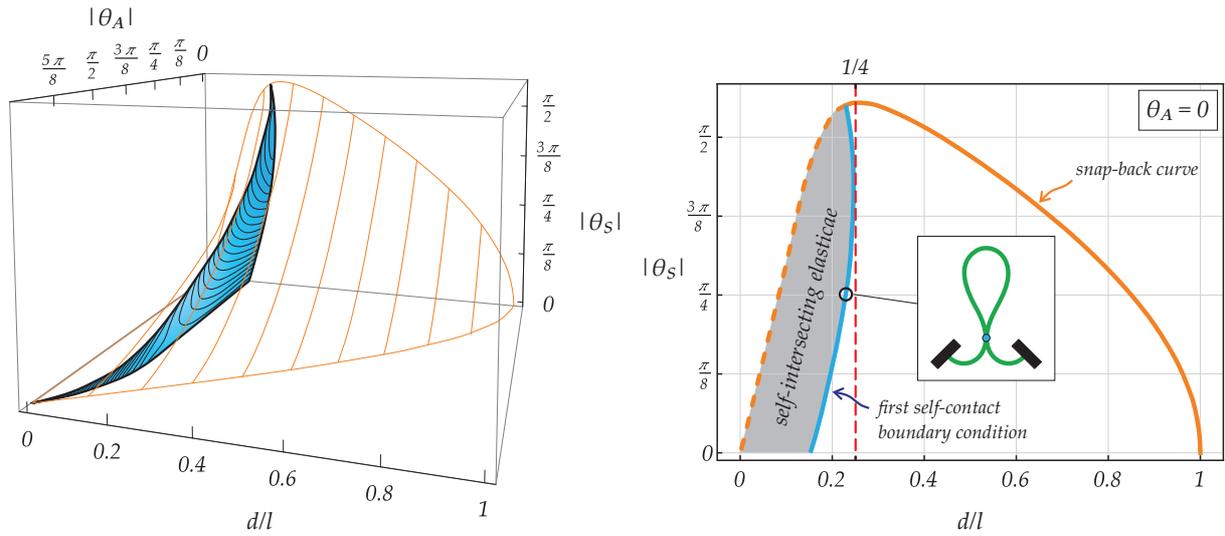}
\end{center}
\caption{\footnotesize (Left) Surface portion for the boundary conditions of first self-contact within the space $d/l-|\theta_A|-|\theta_S|$, which are contained within the snap-back type 1 surface (represented only through its
contour levels). (Right) The most restrictive condition of self-contact, corresponding to the case $\theta_A=0$, is reported as a blue curve within the plane $d/l-|\theta_S|$ showing that
 self-intersecting elasticae may realize before attaining snap-back type 1 (orange curve) only when $d\lesssim l/4$.
 An example of self-intersecting elastica is also reported for $|\theta_S|=\pi/4$.}
\label{fig_selfcomp}
\end{figure}

It is finally observed that self-contact may occur during snapping although the strip has  no self-contact at the critical snap condition.
Indeed, the dynamic transition from the pre and post
snap  configurations may evolve requiring self-intersecting shapes, which could be not geometrically feasible even for Y-shaped strips.

\subsection{Energy release at snapping}

The design of snapping devices can be optimized through the maximization of the energy release during mechanisms.
In a first approximation, the energy release $\Delta \mathcal{E}$ can be estimated as
 the difference in the energy  amounts $\mathcal{E}$, eqn (\ref{energydefinition}), between the
stages before and after the snap-back and evaluated under the quasi-static assumption through eqn (\ref{energyexplicit}).

Restricting attention only to snap-back type 1,
the  elastic energy difference $\Delta \mathcal{E}$
is reported in Fig. \ref{deltaenergysnap1}.
Two  nondimensionalizations are considered, division by $B/l$ (Fig. \ref{deltaenergysnap1}, upper part)
and division by the elastic energy before snapping $\mathcal{E}_0$ (Fig. \ref{deltaenergysnap1}, lower part).
On the left part of Fig. \ref{deltaenergysnap1}, the  elastic energy difference $\Delta \mathcal{E}$  is shown
for different distances
$d/l=\left\{\right.0.05, 0.1, 0.2 ,0.3, 0.4, 0.5, 0.6, 0.7,$ $0.8,$ $0.9$ $\left.\right\}$ for varying the angle for which the snap-back type 1 occurs.
The angle measure is expressed as the modulus of $\theta_A^{sb1}$, the antisymmetric critical angle at the snap-back, normalized through division by $\theta_A^{sb1,bif}$,
the antisymmetric angle at bifurcation for snap-back (the red spots on the $\theta_A$ axis in Fig. \ref{fig_incremdxp06}, \ref{fig_collectiond1} and \ref{fig_collectiond2} and for which no snap
occurs, $\theta_A^{sb1,bif}(d)=\max_{\theta_S} \left\{\theta_A^{sb1}(\theta_S,d)\right\}=\theta_A^{sb1}(\theta_S=0,d)$). For all the reported cases it can be concluded that
the maximum elastic energy release for a fixed distance $d$ is always attained under symmetric conditions, $\theta_A^{sb1}=0$. For such symmetric condition,
the  elastic energy difference $\Delta \mathcal{E}$  is shown
on the right part of Fig. \ref{deltaenergysnap1} as a function of the distance $d$ between the strips' ends. The plot on the upper right part shows that the energy release $\Delta \mathcal{E}$ has a maximum value of
about $76.67 B/l$
at $d\approx0.189 l$ and has null values for both the limit-cases $d=l$ and $d=0$.
The plot on the lower right  part shows that the relative energy release $\Delta \mathcal{E}/\mathcal{E}_0$   monotonically increases with the increase of the distance $d$ and
 attains its maximum ratio of about $0.889$ in the limit condition of $d=l$.
With reference to this last case, it is also worth to highlight 
that the relative energy release  is approximately constant for $d>l/4$, more specifically it varies from $\Delta \mathcal{E}/\mathcal{E}_0(d=l/4)= 0.884$ to $\Delta \mathcal{E}/\mathcal{E}_0(d\rightarrow l)= 0.889$.

\begin{figure}[!htb]
\begin{center}
\includegraphics[width=180mm]{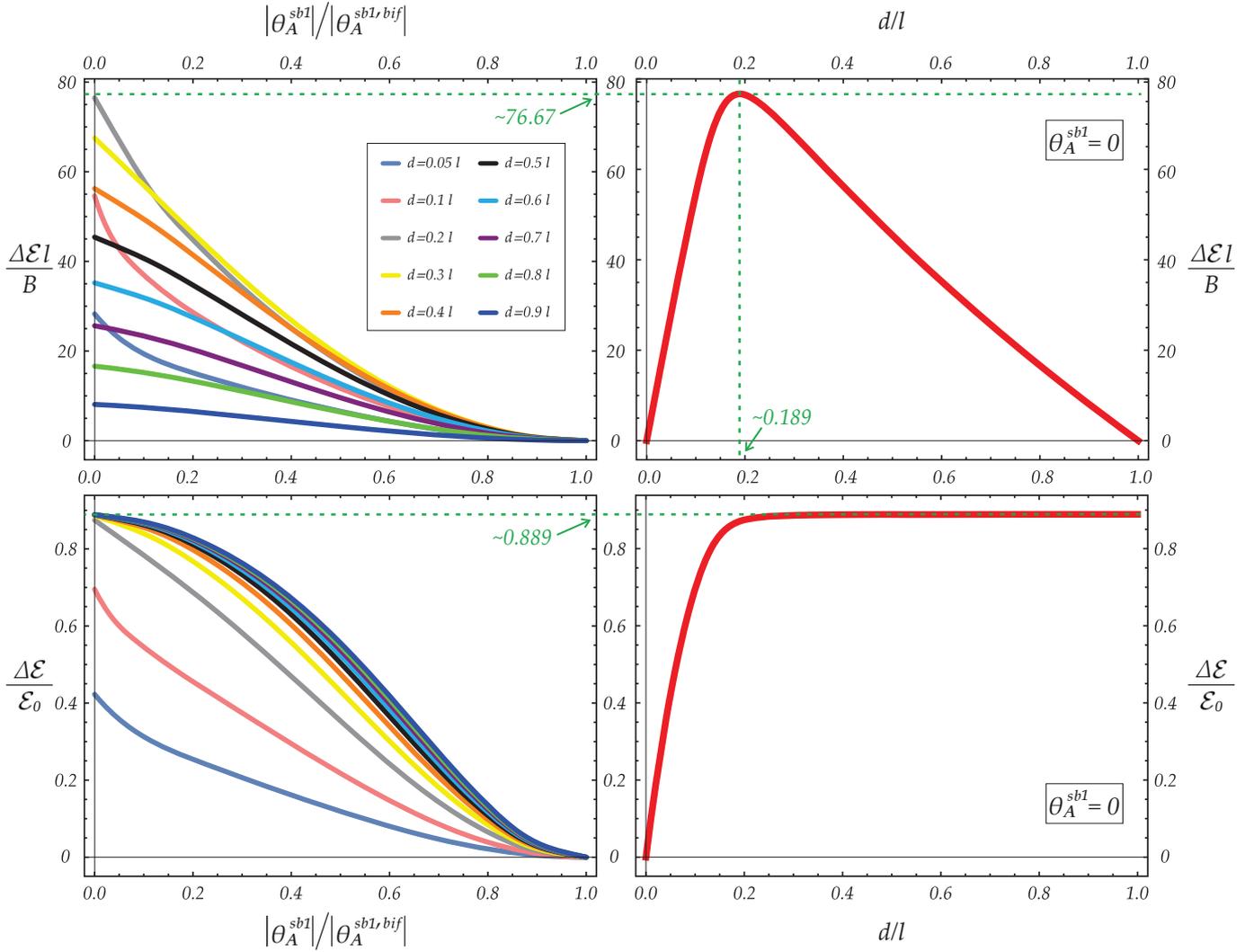}
\end{center}
\caption{
\footnotesize 
Energy release $\Delta \mathcal{E}$ estimated as the difference in the energy amounts $\mathcal{E}$, eqn (\ref{energydefinition}), between the
stages before and after the snap-back of type 1, normalized by $B/l$ (upper part) and by the elastic energy before snapping $\mathcal{E}_0$ (lower part).
The energy release $\Delta \mathcal{E}$ is shown in the left column for different distances
$d/l$ for varying the modulus of the angle for which the snap-back type 1 occurs, $\theta_A^{sb1}$,
normalized through division by $\theta_A^{sb1,bif}=\max_{\theta_S} \left\{\theta_A^{sb1}(\theta_S,d)\right\}$.
With reference to symmetric condition, $\theta_A=0$, the energy release is shown
in the right column as a function of the normalized distance $d/l$.
}
\label{deltaenergysnap1}
\end{figure}

\section{Validation of the analytical predictions}\label{lastsect}
The  universal critical surface for snap type 1 is validated through comparison with experimental observations  and results from numerical simulations. The available experimental data
\cite{beharic},\cite{plautsnap}, restricted to symmetric boundary conditions ($\theta_A=0$), are complemented by testing a  physical model developed to cover
 non-symmetric paths ($\theta_A\neq0$). Finally, the influence of dynamical effects on the system response is assessed through numerical simulation of
 evolutive problems performed in ABAQUS.
  Reliability of the quasi-static predictions obtained through the universal surface is shown in the case
of evolutions with moderate velocity.

\subsection{Experimental results}
A physical model (Fig. \ref{exp}, left) is developed in order to experimentally investigate the snap conditions of the structural system under non-symmetric paths.
Two forks in steel are exploited to control in practice the clamps, constraining the position and the rotation at both
ends of an elastic strip. The strip is obtained from cutting a transparency film by Folex and has cross section 12 mm width times 0.1 mm height and length $l=200$ mm.
Restricting the kinematics of the two forks, specific non-symmetric paths are covered with the developed device.
More specifically, the fork constraining the left end has a fixed position while the fork constraining the right end imposes null inclination, $\theta_l=0$,
and may move  along the $x$-axis.
 From these boundary conditions it follows that the evolution expressed in terms of the
three main kinematic quantities is given by
\beq
d=d(\tau),\qquad
\theta_0=\theta_0(\tau),\qquad
\theta_l=0,
\eeq
so that the non-symmetric paths characterized by $\theta_A(\tau)=-\theta_S(\tau)=\theta_0(\tau)/2$ can be investigated.
Varying only one kinematical parameter, the two following types of experiments are performed:
\begin{enumerate}[Exp. A]
       \item  -- keeping a fixed distance $d(\tau)=\overline{d}$ between the two forks, the rotation at the left end $\theta_0(\tau)$  changes in time;
       \item  -- keeping a fixed rotation at the left end $\theta_0(\tau)=\overline{\theta}_0$, the distance $d(\tau)$  between the two forks changes  in time.
    \end{enumerate}
During each experiment, the rotation $\theta_0(\tau)$ or the distance $d(\tau)$ is slowly varied by hand. The variation in the kinematical parameter is
stopped as soon as the strip snaps and the respective
critical value  is measured for the rotation $\theta_{0,cr}(\overline{d})$ in  Exp. A or for the distance $d_{cr}(\overline{\theta_0})$  in Exp. B with the goniometer or the ruler mounted
on the device, respectively.
The critical conditions experimentally collected from Exp. A and Exp. B are respectively reported as dots  and  crosses within the plane $\theta_0-d/l$ in Fig. \ref{exp} (right)
together with the theoretical critical curve, namely the intersection of the universal snap surface type 1 (Fig. \ref{snap3d}, left) with the plane $\theta_A=-\theta_S$.
\begin{figure}[!htb]
\begin{center}
\includegraphics[width=180mm]{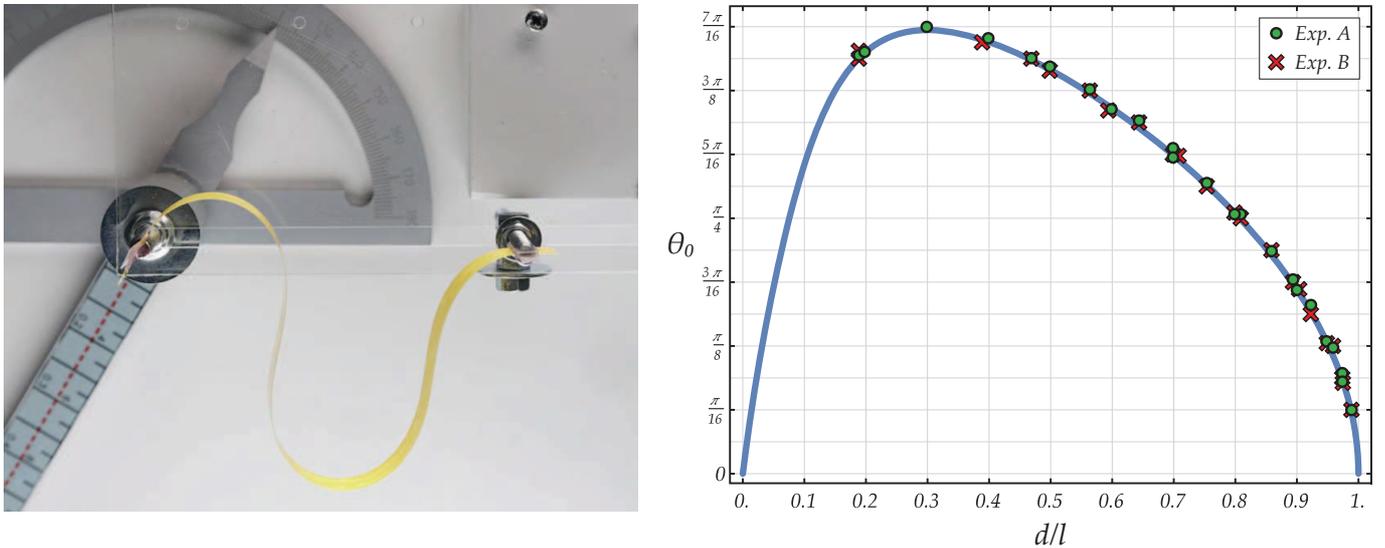}
\end{center}
\caption{\footnotesize (Left) The developed physical model used to experimentally detect the critical snap configurations of the considered system, realized as
a  strip (obtained from cutting a transparency film) constrained at its ends by two forks. (Right) Critical snap conditions $\theta_0-d/l$ from Exp. A (dots) and Exp. B (crosses) confirming the
theoretical predictions (curve) from the present model.
}
\label{exp}
\end{figure}
These experimental measures are also reported together with those measured in the case of purely symmetric rotation $\theta_A=0$  by 
Beharic et al. \cite{beharic} (for $\overline{d}/l\backsimeq\left\{0.833,0.862,0.893,0.926,0.962,0.980,0.990\right\}$)
and by Plaut and Virgin \cite{plautsnap} (for $\overline{d}/l\backsimeq\left\{0.413,0.637,0.827,0.955\right\}$) in Fig. \ref{compare_sections}, where the universal snap surface type 1
and its intersection with planes at constant values of $d/l$, $|\theta_A|$, and $|\theta_S|$ are represented, fully confirming the reliability of the predictions from the present model.
\begin{figure}[!htb]
\begin{center}
\includegraphics[width=180mm]{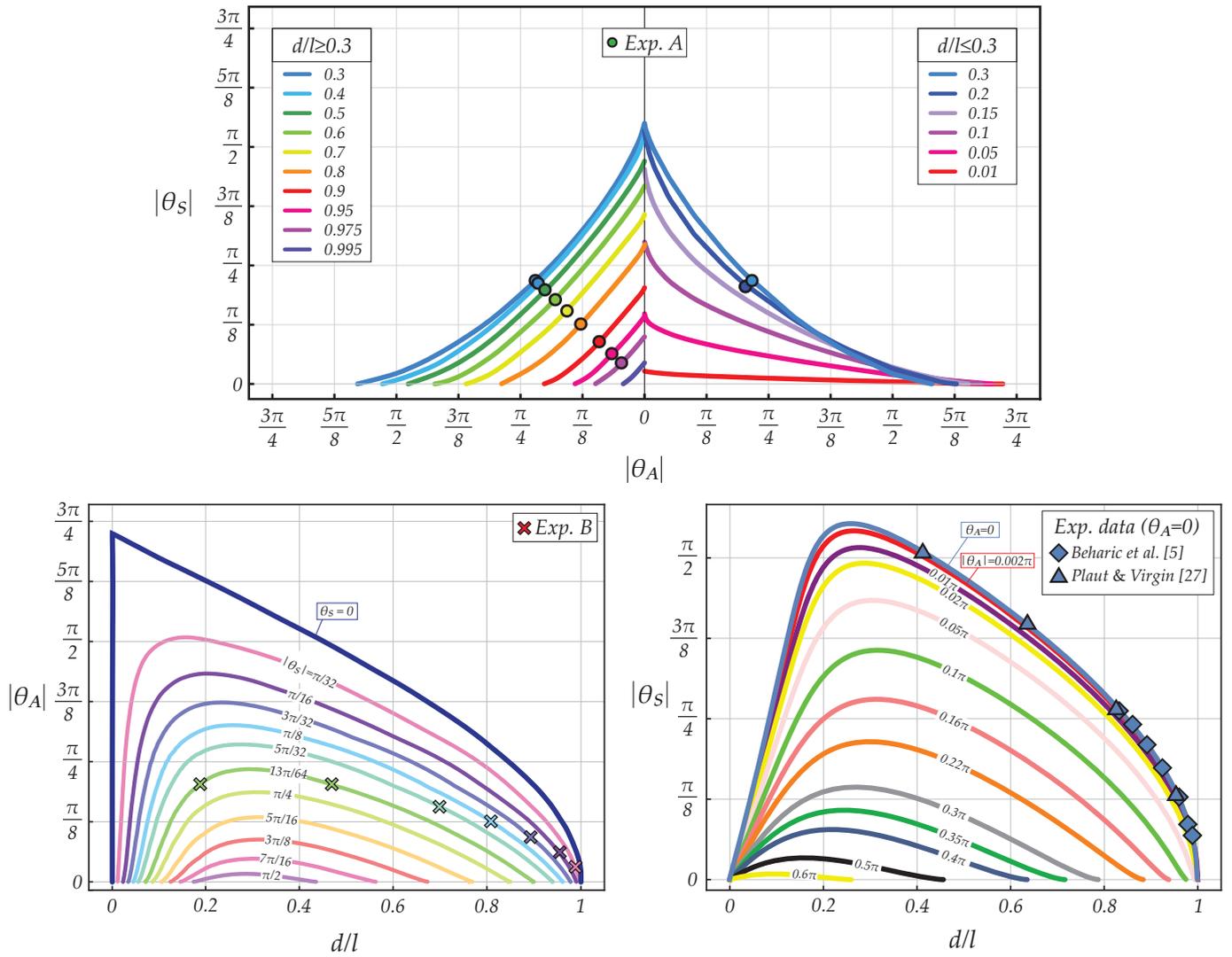}
\end{center}
\caption{\footnotesize Intersections of the universal snap surface type 1 with planes at constant values of $d/l$ (upper part), $|\theta_A|$ (lower part, left), and $|\theta_S|$ (lower part, right).
Critical conditions assessed from Exp. A and Exp. B (performed on the developed physical model, Fig. \ref{exp}, left) confirm the
theoretical predictions reported for fixed $d$ and for fixed $|\theta_A|$, respectively.
Experimental results performed for the purely symmetric case ($\theta_A=0$)
by 
other authors \cite{beharic}, \cite{plautsnap}
 are also reported  (lower part, right),
further confirming the excellent agreement between the experiments and the theoretical predictions from the present model.}
\label{compare_sections}
\end{figure}

\subsection{Numerical simulations and dynamic effects }

In order to evaluate the possible influence of inertia on the snapping conditions of the system,
the presented quasi-static predictions are finally compared with the response obtained from the numerical simulations performed in ABAQUS (v.6.13) for
two different evolutive problems.

\begin{enumerate}[$\mbox{Sim.}$ I]
\item -- Clamps rotating cyclically and with opposite velocity.
Considering a fixed distance $\overline{d}$, the kinematics for the strip's ends is described by
\beq
d=\overline{d},\qquad
\theta_0(\tau)=-\vartheta(\tau)+\overline{\theta}_A,  \qquad
\theta_l(\tau)=\vartheta(\tau)+\overline{\theta}_A,\eeq
or equivalently, through eqn (\ref{thetasymeasym}), in terms of symmetric and antisymmetric parts of the angles as
\beq\label{fixdmotion}
\theta_S(\tau)=\vartheta(\tau),\qquad
\theta_A=\overline{\theta}_A,
\eeq
from which it is evident that the function  $\vartheta(\tau)$ represents the evolution of the symmetric part of the two angles in the
time-like parameter $\tau$ while $\overline{\theta}_A$ represents
a constant anti-symmetric angle during the evolution.
Referring to the physical time $t=T \tau$, where $T=\sqrt{\rho l^4/B}$ is the  characteristic time  for the system having a linear mass density $\rho$,
results from Sim. I are reported for
a fixed antisymmetric rotation $\overline{\theta}_A=0.16\pi$ and two fixed distances $\overline{d}/l= 0.4$ (Fig. \ref{snapex1}, upper part, left) and $\overline{d}/l= 0.8$
(Fig. \ref{snapex1}, lower part, left).
In both cases, the cyclic evolution in the boundary conditions is realized through
the succession of the increase and decrease in the symmetric rotation within the range $\theta_S\in[-\pi/2,\pi/2]$
keeping a constant modulus in the velocity,
\beq
\left|\dot{\vartheta}(\tau=t/T)\right|=\frac{\Omega}{T},
\eeq
where a superimposed dot corresponds to the derivative in the physical time $t$ and
$\Omega$ is the dimensionless (angular) velocity.
The cyclic path prescribed in Sim. I theoretically encounters two snap conditions given by the same modulus of $\theta_S$.

\item -- Monotonic variation  in the clamps distance.
Constant rotations $\overline{\theta}_0$ and $\overline{\theta}_l$ are assumed for the two ends, so that the kinematical evolution of the strip's ends is given by
\beq
d=d(\tau),\qquad
\theta_0=\overline{\theta}_0,  \qquad
\theta_l=\overline{\theta}_l,\eeq
where the function $d(\tau)$ defines a monotonic increase or decrease in the  distance between the two clamps.
Results from Sim. II are reported for fixed rotations $\overline{\theta}_0$ and  $\overline{\theta}_l$ such that $\overline{\theta}_S= \pi/8$ (Fig. \ref{snapex1}, upper part, right) and $\overline{\theta}_S= \pi/4$
(Fig. \ref{snapex1}, lower part, right), and in both cases $\overline{\theta}_A=0.16\pi$.
In both cases, the monotonic variation in the distance $d$ is considered from the value $d(\tau=0)/l= 0.336$
with a constant  velocity,
\beq
\left|\dot{d}(\tau=t/T)\right|=\frac{\Delta \,l}{T},
\eeq
where $\Delta$ is the dimensionless velocity. During both the monotonic shortening and the monotonic lengthening
in the clamps distance prescribed in Sim. II, a snap mechanism is theoretically predicted for each evolution.
 \end{enumerate}

A linear viscous Rayleigh damping acting on the mechanical system,  modeled as 100 planar beam elements with linear elastic
constitutive behaviour, is considered in all the simulations through the mass-proportional and the stiffness-proportional damping coefficients
respectively as $A_d=8.25 \times 10^{-3} /T$  and $B_d=6.06 \times 10^{-3} T$. The inherent extensibility of the strip modeled in ABAQUS has been considered through the  axial stiffness $EA=10^6 \times B / l^2$.
All the presented analyses are performed using the nonlinear geometry option and started from the undeformed straight configuration with null rotations, $d=l$ and $\theta_0=\theta_l=0$.
All the simulations share the first two static steps, while are different in the last dynamic step as follows:
\begin{enumerate}[$\mbox{Step}$ 1]
\item -- Static: An end's distance $d(0)$ is imposed and
a transversal load is applied in order to achieve the buckled configuration;
\item -- Static: the transversal load is removed and the clamp rotations are imposed in order to set the initial values of $\overline{\theta}_A$ and $\theta_S(0)$;
\item -- Dynamic implicit: inertial effects are analyzed during the evolution in the boundary conditions at velocity with constant modulus from $t=0$ to $t=t_{f}$ as
\begin{enumerate}[$\mbox{Sim.}$ I]
\item -- initial and final configurations have $\theta_S(0)=\vartheta(0)=\vartheta(t_{f})=0$. Introducing the reference time
$T_{r}=\pi T/(2\Omega)$, the duration of the evolution is given by $t_f=4 T_r$. The rotation velocity is assumed $\dot{\vartheta}=\Omega/T$  for $t\in[0,T_{r}]$ and $t\in[3T_{r}, 4 T_r]$, while is assumed
$\dot{\vartheta}=-\Omega/T$ as for
$t\in[T_{r}, 3T_r]$, so that the velocity changes sign whenever the modulus of rotation reaches $|\theta_S|=\pi/2$;
\item -- the initial and final configurations have the ends distance given by $d(0)$ and $d(t_{f})$. The transition between these two conditions
occurs with constant velocity and has a duration $t_f=T_r$, where $T_{r}=0.637 T/\Delta$. The constant velocity is taken positive or negative in order to investigate the
snap mechanism during lengthening or shortening.
\end{enumerate}
\end{enumerate}

The results of the simulations of the two evolutive problems are reported in Fig. \ref{snapex1}  and compared with the respective quasi-static behaviour predictions using the inextensible elastica (and highlighting the snap angles and snap distances, defined by the snap curve $\theta_A=0.16\pi$  in Fig.\ref{compare_sections} right, lower part).
In particular, results for the Sim. I are reported in Fig. \ref{snapex1} (left) in terms of the moment (at the right end) at cyclically varying (symmetric) rotation $\theta_S$, while those for the second evolutive problem
are reported in Fig. \ref{snapex1} (right) in terms of the horizontal reaction (at the right end) for monotonic increase and decrease in the clamps distance.
The respective numerical results are reported for three values of dimensionless velocities, $\Omega=\left\{0.01,\,0.1,\,1\right\}$ and  $\Delta=\left\{0.001,\,0.01,\,0.1\right\}$,
 showing that the quasi-static model accurately describes the mechanical behaviour of the structural system until approaching the snap-back conditions,
identified as the intersection of the loading path with presented snap-back surface type 1. Due to the presence of dissipative effects,
the post-snap quasi-static path is reached after a transient time from the snap for which the dynamical effects are decayed.
More specifically,  non-negligible dynamic effects lead to a delay in the occurrence of snap with respect to the quasi-static prediction for high velocities.
Oppositely, the dynamic response becomes almost completely
 superimposed (except for a small transient) to the quasi-static curve in the case of a velocity $\Omega=0.01$ and $\Delta=0.001$, values defining
 the velocity orders below which the present model, although obtained within
 the quasi-static framework, fully represents a reliable model.

\begin{figure}[!htb]
\begin{center}
\includegraphics[width=180mm]{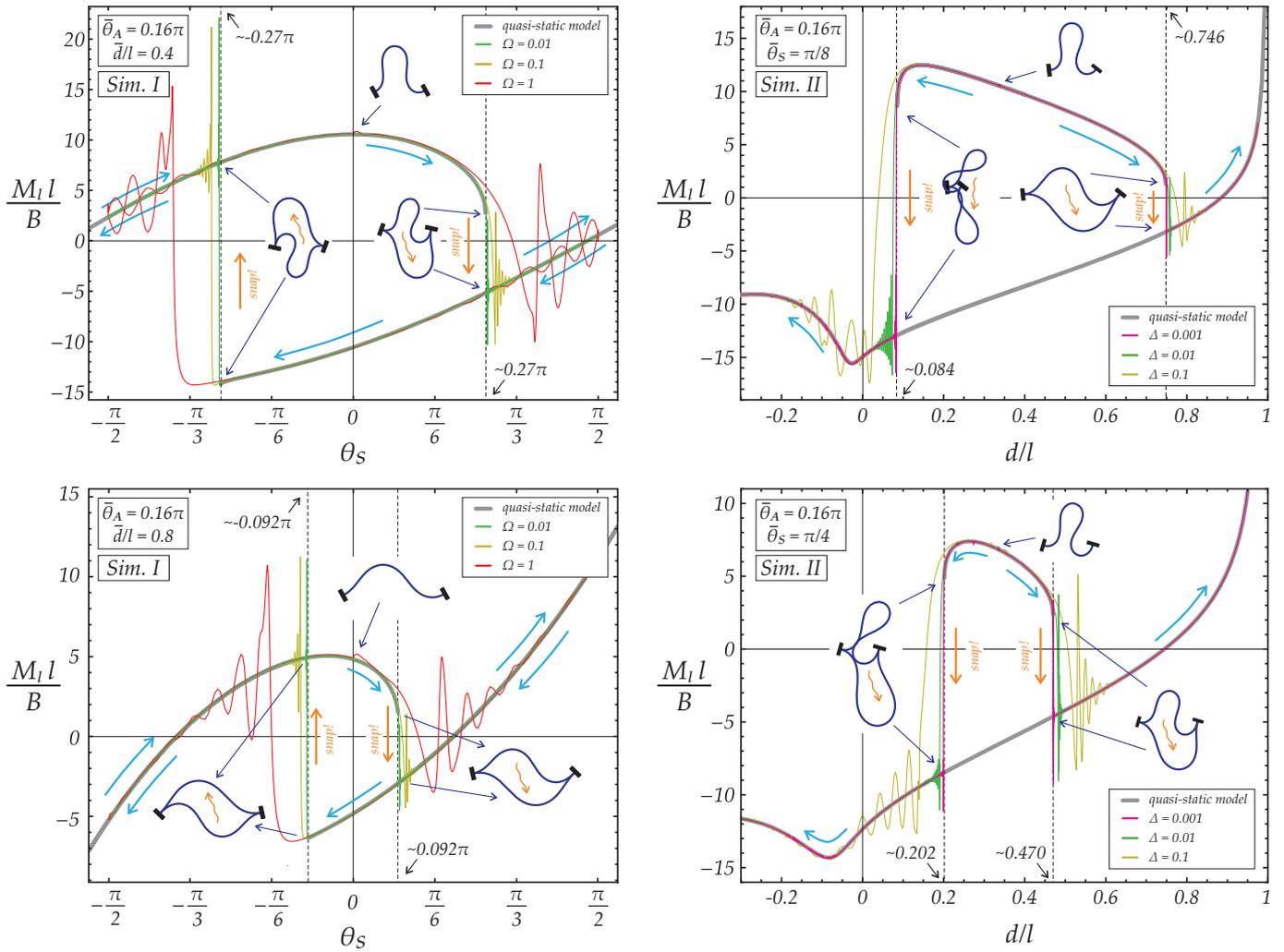}
\end{center}
\caption{
\footnotesize 
Theoretical predictions from the quasi-static model versus results from numerical
simulations performed through finite element code (ABAQUS) for different values of the dimensionless velocities $\Omega$ and $\Delta$.
(Left column) $M_l-\theta_S$ response from Sim. I for a cyclically varying symmetric rotation $\theta_S$ at fixed antisymmetric rotation $\overline{\theta}_A=0.16\pi$
and fixed clamps distance
$\overline{d}/l=0.4$ (upper part) and $\overline{d}/l=0.8$ (lower part). 
(Right column) $M_l-d/l$ response from Sim. II for a monotonic increase and decrease in the distance $d$ from $d(\tau=0)/l=0.336$,
with fixed antisymmetric rotation $\overline{\theta}_A=0.16\pi$ and fixed symmetric rotations $\overline{\theta}_S=\pi/8$ (upper part) and $\overline{\theta}_S=\pi/4$ (lower part). 
}
\label{snapex1}
\end{figure}

\section{Conclusions}
Within a quasi-static framework, the number of stable equilibrium configurations has been disclosed for an elastic strip
for varying the parameters controlling the kinematics of its  ends. This analysis has led to the definition of universal snap surfaces,
collecting the critical values of ends distance and rotations for which the strip shows a snap mechanism. Available experimental data and
experimental results  from testing a developed physical model fully validate the presented theoretical universal snap surface. Finally, finite  element simulations show the influence of inertia on the snapping mechanisms and, in the case when the controlled ends move moderately, confirm the theoretical predictions based on the present quasi-static  model.
These results are complemented by the dimensionless analysis of elastic energy release at snapping, towards the optimal design of impulsive motion.
In addition to the relevant contribution to the stability of structures, the present results may find application
in a wide range of technological fields, ranging from energy harvesting to jumping robots.

\vspace*{5mm} \noindent {\sl Acknowledgments} The authors gratefully
acknowledge financial support from the ERC advanced grant \lq
Instabilities and nonlocal multiscale modelling of materials'
ERC-2013-AdG-340561-INSTABILITIES (2014-2019). 

\end{document}